\documentclass{aa} 

\newcommand{\swift}{\textit{Swift}}

\newcommand{\xmm}{\textit{XMM-Newton}}
\newcommand{\XMM}{\textit{XMM-Newton}}
\newcommand{\nustar}{\texttt{}{NuSTAR}}

 \usepackage{tablefootnote}
  \usepackage{xcolor}
  \usepackage{natbib}
  \usepackage{graphicx}
  \usepackage{lscape}
  \usepackage{multirow}
  \usepackage[normalem]{ulem}
  \usepackage{multirow}
  \usepackage{txfonts}
  \usepackage{geometry}
  
\begin{document}

     \title{The rise and decline of a transient obscuration event in a moderately distant type-I quasar}

     \titlerunning{The rise and decline of a transient obscuration event}
      \authorrunning{A. Akylas et al.}
     
     \author{A. Akylas
            \inst{1} 
            \and
             A. Georgakakis
            \inst{1}
            \and
             E. Pouliasis
            \inst{1}
            \and
            I. Georgantopoulos 
            \inst{1}
            \and 
            C. Ricci 
            \inst{2,3}
            \and 
            M. Chira 
            \inst{1}
            \and
            A. Ruiz
            \inst{1}
            \and
            E. M. Xilouris
            \inst{1}
                        }
\institute{
$^{1}$ Institute for Astronomy Astrophysics Space Applications and Remote Sensing (IAASARS), 
National Observatory of Athens, Ioannou Metaxa \& Vasileos Pavlou, Penteli, 15236, Greece \\
$^{2}$ Department of Astronomy, University of Geneva, Chemin d'Ecogia 16, 1290 Versoix, Switzerland \\
$^{3}$ Instituto de Estudios Astrof\'isicos, Facultad de Ingenier\'ia y Ciencias, Universidad Diego Portales, Av. Ej\'ercito Libertador 441, Santiago, Chile
}

\date{}        

 \abstract
    {Outflows of ionised material, often dubbed warm absorbers (WAs), are common in nearby Seyferts but sparse in distant and more energetic quasars (QSOs). This paper presents multi‑epoch X‑ray and optical observations of the serendipitously identified type-I QSO 2MASS\,J14302580+4159572 at a moderate redshift of $z = 0.35$, which trace the rare case of the rise and decline of a transient ionised absorber over a 22-year timescale.
   The discovery observations were carried out by \XMM\ between 2002 and 2005. The first two (2002 and 2003) show unremarkable X-ray spectra typical of unobscured type-I QSOs. The third observation in 2005, however, reveals a pronounced suppression of the X-ray spectrum at energies below  $\sim3\,\rm keV$. Spectral analysis demonstrates that the 2005 event is produced by a moderately ionised absorber with column density $N_H\simeq5\times10^{22}\,\rm cm^{-2}$ and ionisation parameter $\rm log\,\xi\simeq1.4$. Simultaneous X-ray data analysis of \xmm, \nustar, and \swift/XRT data over a 22‑year baseline indicates  that the absorber remains stable for nearly 18 years before weakening  in the \swift/XRT 2023-2024 observations. Throughout this period, the intrinsic 3--10\,keV continuum remains remarkably stable, confirming that the variability is driven by evolving line‑of‑sight ionised absorption rather than intrinsic luminosity fluctuations.  We interpret the event as a thermally driven ionised outflow located outside the broad-line region (BLR) at sub‑parsec to parsec distances with an estimated density of about $10^{6}\,\mathrm{cm^{-3}}$.
   Corroborating this picture, the high-ionisation [O\,III]~$\lambda4363$ line in the optical spectra vary in tandem with the X-ray absorption, suggesting that the outflow progressively modifies the ionisation state of the inner narrow-line region (NLR) on decadal timescales.}

    \keywords{X-rays: general -- galaxies: active -- quasars: general -- quasars: supermassive black holes}

\maketitle

\section{Introduction}

Active galactic nuclei (AGNs) are powered by the accretion of matter onto supermassive black holes (SMBHs) that exist at the centres of galaxies. During this process, matter, mostly in gaseous form, experiences viscous and magnetic stresses that cause it to lose angular momentum and spiral inwards, forming an accretion disc around the central compact object. The dissipation of these internal stresses heats the disc, efficiently converting gravitational potential energy into radiation \citep[e.g.][]{Frank2002}. The resulting radiation field can be sufficiently intense to launch outflows that interact with matter on larger scales giving rise to observable signatures such as absorption or emission spectral features \citep[e.g.][]{king2003,crenshaw2003}. AGN outflows exhibit a broad phenomenology,  including highly ionised and fast moving gas observed as absorption troughs against the X-ray and/or UV/optical continuum (e.g. ultra-fast outflows), weakly ionised material producing extended UV/optical emission line profiles, and neutral gas observed in emission at millimetre wavelengths. However, these phenomena may represent components of a single, large-scale, stratified wind observed at different distances from the SMBH \citep{tombesi2013, laha2021}. Previous studies have also shown that such outflows can mediate the interaction between SMBHs and their host galaxies, potentially explaining the $M_{\mathrm{BH}}-\sigma$ and $M_{\mathrm{BH}}-M_{\mathrm{bulge}}$ relations \citep[e.g.][]{blustin2005}.

One of the most common manifestations of AGN winds is warm absorbers (WAs), which represent outwards expanding ionised material. They were first identified in the  X-ray spectra of nearby Seyferts as deep photoelectric absorption edges from O\,\textsc{vii} and O\,\textsc{viii} \citep[e.g.][]{reynolds1997,george1998}. Such features, dominated by K-shell ions of the lighter metals (e.g. C, N, O, $\sim0.3$--$0.7\,\mathrm{keV}$) and by L-shell ions of iron ($\sim0.7$--$1.2\,\mathrm{keV}$) \citep[e.g.][]{blustin2005} are found in nearly 50\% of local AGNs \citep[e.g.][]{reynolds1997,george1998} with outflow velocities of a few hundred to a few thousand $\rm km\,s^{-1}$. The ionisation parameter, defined as the ratio of the density of ionising photons to the density of the gas, spans nearly four orders of magnitude, $\xi\approx 10^{-1}-10^{3}\,\rm erg\,cm\,s^{-1}$, and the column densities cover three orders of magnitude, $N_{\mathrm{H}} \approx 10^{20}-10^{23}~\mathrm{cm^{-2}}$ \citep[e.g.][]{blustin2005,mckernan2007,laha2014}.

Observational studies show that WAs are not continuously present. Their appearance, disappearance, and long-term variability indicate that they are episodic phenomena, detectable only during specific phases of AGN activity \citep[e.g.][]{blustin2005,krongold2007,crenshaw2012,kaastra2014}. Although they carry only a small fraction of the AGN bolometric power, the total mass processed through AGN outflows over the lifetime of the system is likely sufficient to deliver significant feedback to their environments \citep[e.g.][]{blustin2005,crenshaw2012}. Detecting WAs beyond the local Universe ($z\gtrsim0.1$) is challenging because their characteristic soft X-ray absorption lines are redshifted outside the spectral window of X-ray detectors, and distant AGNs are often too faint to provide the high signal-to-noise spectra required to identify these features. Recently, \citet{waddell2024} presented statistical evidence for the presence of WAs in 12\% of their hard X-ray–selected sample from the eROSITA Final Equatorial Depth Survey field; the sources have a median redshift of $z = 0.35$.   However, in contrast to the extensive monitoring of these phenomena in the nearby Universe, similar multi-epoch X-ray studies of ionised absorption in more distant quasars (QSOs) remain rare. A characteristic example is the work of \citet{Markowitz2024}, who used the timing capabilities of  eROSITA to track month-scale changes in obscuration in the type-I AGNs EC\,04570$-$5206 at $z=0.276$.

In this work, we analyse multi-epoch X-ray and optical observations of the QSO 2MASS\,J14302580+4159572 at a redshift of $z = 0.35$. The source was serendipitously identified while mining the \xmm\ archive for QSOs in the Sloan Digital Sky Survey Quasar Catalog Data Release 16 \citep[SDSS DR16,][]{Higley2020} that exhibit significant spectral variability. We find evidence for variable ionised absorption in the X-ray spectra of 2MASS\,J14302580+4159572, manifested by a strong modulation of the continuum at soft energies ($\lessapprox 3\, \rm keV$) over a timescale of about 22 years. We argue that the multi-epoch data of 2MASS\,J14302580+4159572 trace a rare case of the rise and decline of a transient ionised obscuration event in a type-I QSO, possibly associated with an outflow.  
The structure of this paper is as follows: in Sections~2 and 3, we describe the X-ray observations and the UV/optical data, respectively. Section~4 presents the X-ray data reduction. In Section~5, we introduce the X-ray spectral modelling, in Section~6 we discuss the behaviour of the optical/UV photometry, and in Section 7 we present the optical spectra of the source. In Section 8 we present basic geometrical models for the transient event explanation. The discussion and summary are provided in Sections 9 and 10, respectively.

\section{X-ray observations}\label{xrayobservations}

2MASS\,J14302580+4159572 is serendipitously detected on three \xmm\ observations lying  at an angular separation of $\sim$4.5\,arcmin from the primary target (GB1428+4217). The European Photon Imaging Camera \citep[EPIC;][]{Struder2001} instruments were operated in full-frame mode with the thin filter applied.  The Reflection Grating Spectrometer \citep[RGS;][]{Herder2001}, configured in spectroscopy HER+SES mode, was active during the observations but does not yield usable data because of the faintness of the source.  

The source also lies within the footprint of two \nustar\ archival observations, at angular separations of 4.3 and 4.8\,arcmin from their respective targets, GB1428+4217 and 5BZQJ1430+4204. Cross-matching the position of 2MASS\,J14302580+4159572 with the Neil Gehrels  \swift\ Observatory X-Ray Telescope (XRT) Living Point Source Catalogue (LSXPS), maintained by the UK \swift\ Science Data Centre \citep{evans2009}, yields a counterpart within 2.9\,arcsecond, identified as LSXPSJ143025.6+415959. LSXPS includes both individual \swift/XRT observations and stacked images that allow the recovery of faint sources by combining data from multiple epochs. 2MASS\,J14302580+4159572 lies in 68 \swift/XRT datasets observed between 2008-08-24 and 2024-05-09. The total duration of these observations is about 70\,ks. The complete X-ray observing log is provided in Table~\ref{xrayobservations_log}.

\begin{table}
\caption{X-ray observation log of 2MASS\,J14302580+4159572.}\label{xrayobservations_log}
\begin{tabular}{lccc}
\hline
Telescope & ObsID & Date  & Exp. Time (ks)  \\
\hline
\multirow{3}{*}{\xmm}    & 0111260101  &  2002-12-09 & 18.9 \\
                        & 0111260701  &  2003-01-17 &  14.6 \\
                        & 0212480701  &  2005-06-05 &  19.7 \\
\multirow{2}{*}{\nustar} & 60001103002 &  2014-07-14 & 49.2 \\
                        & 90901634002 &  2023-12-09 &  58.7 \\
\texttt{\swift/XRT}     & 68 datasets &  2008--2024 & $\simeq70$   \\
\hline
\end{tabular}
\tablefoot{
The table lists the identification numbers (ObsID) of the X-ray observations of the source carried out by the different telescopes along with the corresponding date and exposure time in kilo-seconds. For \swift/XRT, this time represents the cumulative exposure of all datasets.
}
\end{table}

\section{UV/optical observations}\label{opticalobservations}

{The Optical Monitor (OM) onboard \xmm\, provides UV/optical photometry which is simultaneous to the X-ray data.  However, because of the smaller field of view of the OM compared to the EPIC instruments, only the observation 0111260101 (see Table \ref{xrayobservations_log}) provides photometric measurements for the source in the B and UVW1 filters. The source also lies outside the footprint of the \swift\, UVOT \citep[Ultraviolet/Optical Telescope;][]{Roming2005}. This is likely because of  differences in the field of view sizes of \swift/XRT and UVOT and/or different observing modes of the two instruments during the acquisition of the data. 

Archival optical data are also available from large ground-based surveys. The Sloan Digital Sky Survey Data Release 7 \citep[SDSS DR7;][]{sdss_dr7} provides photometric observations ($ugriz$ bands) of 2MASS\,J14302580+4159572 obtained on 2003 January 27th. The Legacy Survey Data Release 10 \citep[LS10;][]{legacy2019} provides co-added optical photometry for the source from observations taken in the period between 2016~February~26th and 2018~February~25th.
Optical spectra of 2MASS\,J14302580+4159572 are available from SDSS (observed on 2003~March~25th) and the Dark Energy Spectroscopic Instrument Data Release 1 \citep[DESI DR1;][]{des-dr1}, observed on 2022 January 13th. 

These archival datasets are complemented with higher-cadence optical monitoring from the Zwicky Transient Facility \citep[ZTF;][]{Bellm2019,Masci2019}. Photometry in the $gri$ bands is extracted through the ZTF Forced Photometry Service \citep{Masci2023} at the SDSS optical QSO position. We filter out epochs with ZTF quality flags \texttt{procstatus>0} and \texttt{infobitssci>0} to remove observations affected by non-photometric conditions and/or poor photometric calibration. We further remove potentially problematic measurements using the ZTF \texttt{scisigpix} metric, which provides a robust estimate of the per-pixel standard deviation. An iterative sigma-clipping approach is adopted, whereby we exclude epochs with \texttt{scisigpix} values more than 2.5 standard deviations from the mean. For the $g$ and $r$ bands, these quality cuts yield a total of 936 and 817 photometric epochs, respectively, between March~2018 and July~2025. We disregard the $i$ band light curve because of the lower number of epochs (209), the sparser and more inhomogeneous cadence compared to the $gr$ filters, and the overall lower signal-to-noise per observation. For brevity, Fig.~\ref{fig:lc_ztf} shows only the $g$ band absolute photometry light curve, generated following the methods described in \citet{Masci2023}.
Fig. \ref{fig:lc_master} visualises the multi-wavelength observation epochs of 2MASS\,J14302580+4159572. The upper plot summarises the available X-ray data from \xmm\ and \swift/XRT (see Table \ref{xrayobservations_log}). The lower panel summarises the optical photometry from ZTF, SDSS, LS10 and OM. In the same figure we also indicate the epochs of the available optical spectroscopy from SDSS and DESI, as well as the dates of the \nustar\ observations.  
 
\begin{figure}
	\includegraphics[width=1\columnwidth]{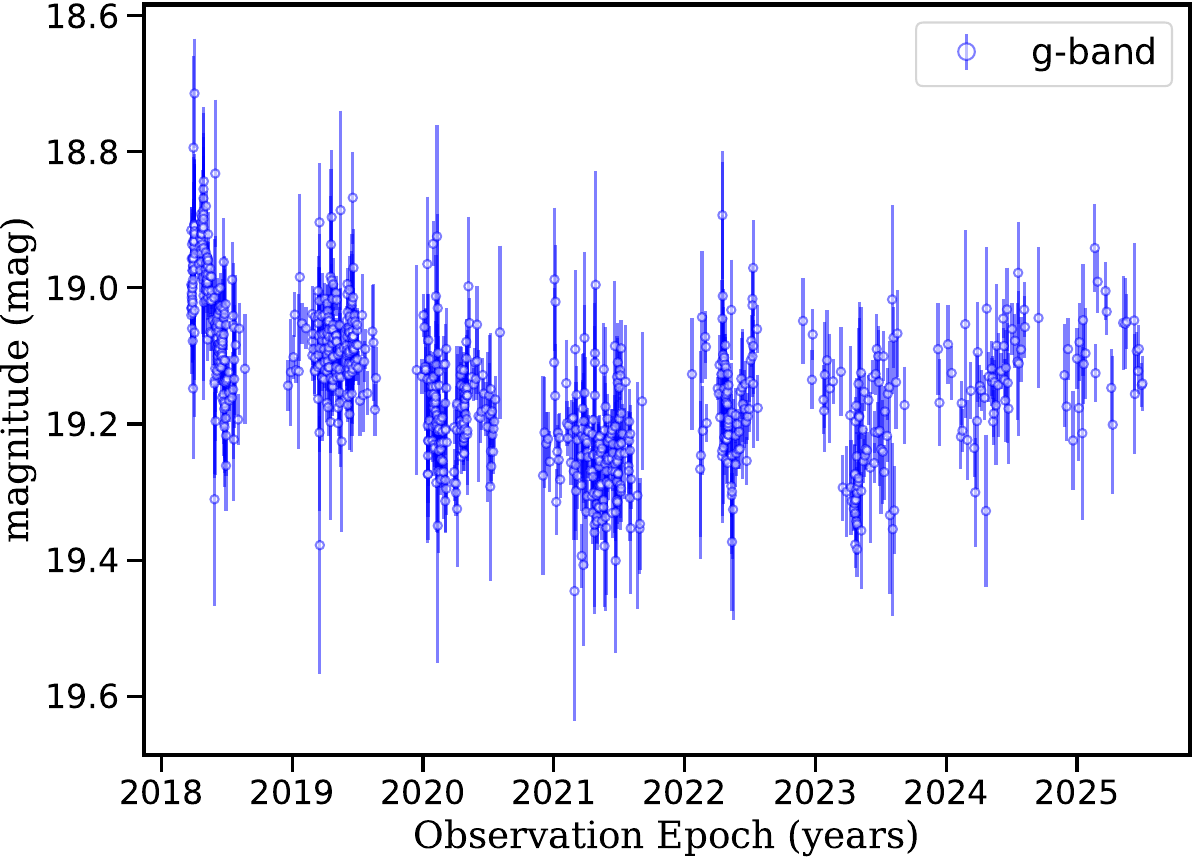}
    \caption{ZTF light curve of 2MASS\,J14302580+4159572 in the $g$ band. The light curve shows the ZTF $g$ band photometry of the source
    (see text for details). The ZTF $r$ band photometry follows a very similar temporal pattern and is not shown for brevity. 
    } 
    \label{fig:lc_ztf}
\end{figure}

\begin{figure*}
	\includegraphics[width=1\textwidth]{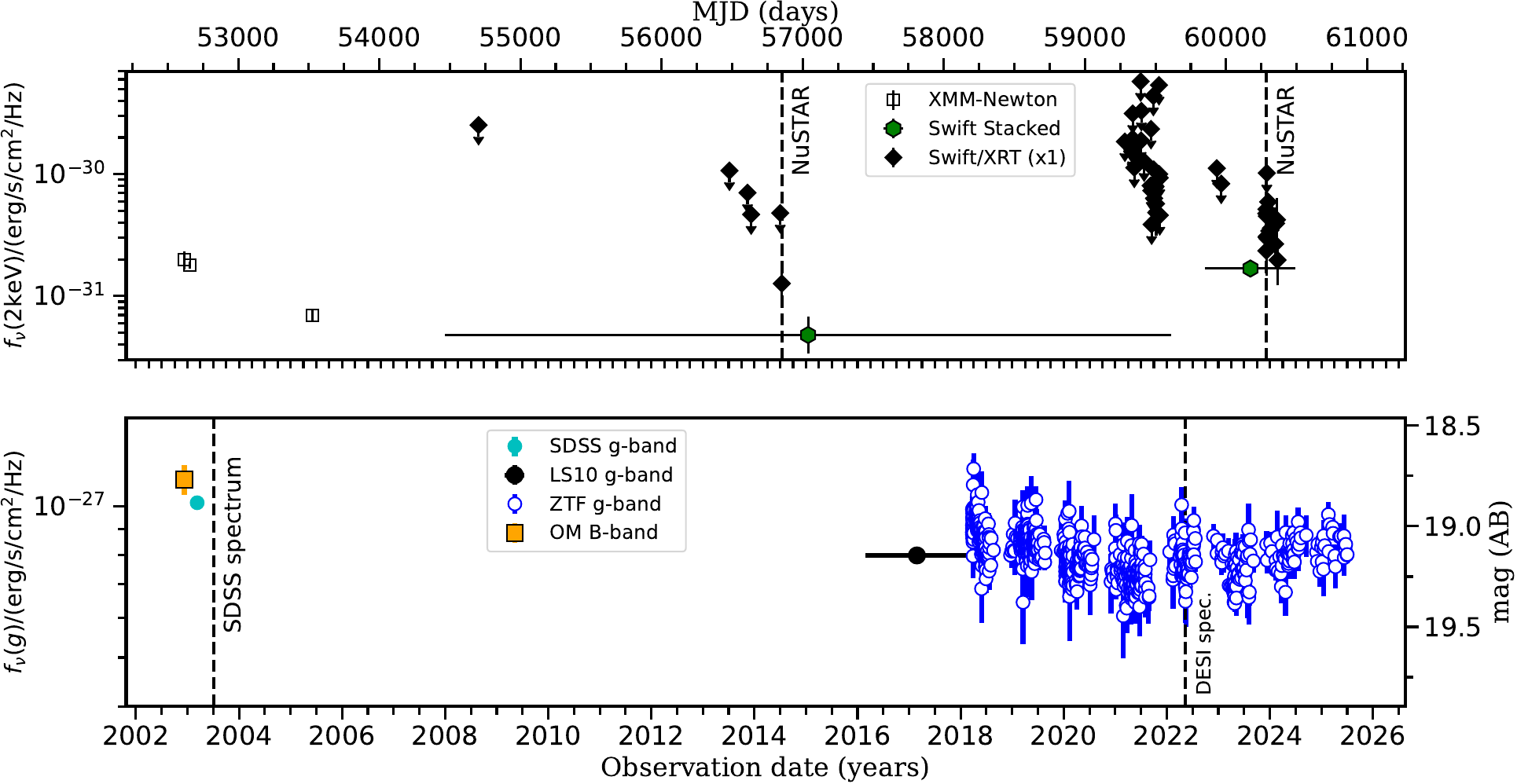}
    \caption{Observation epochs of multi-wavelength datasets for 2MASS\,J14302580+4159572. The top panel shows the X-ray flux density at 2\,keV  for the \XMM\ (black open squares), \swift/XRT individual observations (black filled diamonds, upper limits are shown with an arrow pointing down) and stacked \swift/XRT observations (green hexagons; see text for details). The observed 2\,keV flux density for \swift/XRT individual observations is estimated from the observed flux in the 0.5-2\,keV band assuming a simple X-ray power-law spectral model with photon index $\Gamma=1.9$ \cite[i.e. similar to unobscured QSOs;][]{nandra1994}. Adopting instead a flatter index, $\Gamma=1.0$, similar to the \XMM\ 2005 epoch (see Section \ref{SpectralAnalysis})  shifts the data points of the Swift/XRT individual observations by 0.3\,dex towards brighter flux densities. The  2\,keV flux density for the \XMM\ and \swift/XRT stacked data is estimated from the spectral fits of Section \ref{SpectralAnalysis}. The epochs of the \nustar\ observations (see Table \ref{xrayobservations_log}) are also indicated in the plot. The bottom panel shows the optical $g$ band flux density from ZTF (blue open circles, see also Fig. \ref{fig:lc_ztf}), SDSS (cyan filled circle), Optical Monitor $B$ band (orange square; $B$-band magnitude is converted to $g$ band using the transformations of \citealp{Jester2005}) and LS10 (filled black circle). The epochs of the SDSS and DESI spectroscopy are also indicated with the vertical dashed lines. For convenience the right-hand side y-axis of the bottom panel shows the scaling between flux density and AB magnitude for the ZTF, SDSS, LS10 and OM optical datasets.} 
    \label{fig:lc_master}
\end{figure*}

\section{X-ray data reduction}\label{datareduction}

\subsection{\xmm}

We analyse \xmm\ EPIC data from both the PN and MOS detectors using the Pipeline Processing System (PPS) calibrated event lists from the \xmm\ Science Archive. The PPS observations are further processed using the \xmm\ Science Analysis Software ({\sc sas}\,v21.0.0). The event files are screened for high particle background periods. For each observation, we extract a high-energy (10--12\,keV), single-pixel (PATTERN=0) light curve to visually identify periods of elevated background. Threshold count rates of about $\rm 0.5\,counts\,s^{-1}$ for PN and $\rm 0.25\,counts\,s^{-1}$ for MOS are defined to isolate intervals of low and stable background. These thresholds are used to generate Good Time Interval (GTI) files via the {\sc tabgtigen} task of {\sc sas}, which are then applied during the data selection process via the {\sc evselect} task of {\sc sas}. 

For the spectral analysis we select standard patterns for the X-ray events, singles and doubles for the EPIC PN (i.e. PATTERN 0--4) and singles, doubles, triples, and quadruples for EPIC MOS (PATTERN 0--12). Additionally, we apply the FLAG=0 criterion within the {\sc evselect} task of {\sc sas} to exclude events occurring near CCD gaps, outside the EPIC field of view and/or affected by bad pixels.

For the spectral extraction we define a circular region with a radius of 15\,arcsec centred on the source. A nearby source-free region with a radius of 50\,arcsec is used for background estimation. Source and background event files were generated using the {\sc evselect} task of {\sc sas}. The corresponding response matrix files (RMFs) and ancillary response files (ARFs) are produced using the {\sc rmfgen} and {\sc arfgen} tasks of {\sc sas}, respectively. All spectra are grouped to ensure a minimum of 15 counts per bin, enabling statistically reliable $\chi^2$ analysis.

\subsection{\nustar}

The \nustar\ observations are processed using the data analysis software {\sc NuSTARDAS} v2.1.2 and the calibration database v.20231017\footnote{\url{https://heasarc.gsfc.nasa.gov/docs/nustar/analysis/nustar\_swguide.pdf}}.  
We download the calibrated event list files from the \nustar\ archive. Then, we extract the source and background event files for each of the two \nustar\ focal plane modules (FPMA \& FPMB) using the {\sc nuproducts} script. We  adopt a radius of 50" for the source spectral extraction, for both FPMA and FPMB. The background  spectra are extracted from a four times larger (100" radius) source-free region of the image at an off-axis angle similar to that of the source. 

\subsection{\swift/XRT}

In the case of the \swift/XRT data, we do not perform the X-ray data reduction manually. The X-ray spectral data are generated automatically using the \swift/XRT Science Data Centre repository web-form\footnote{\url{https://www.swift.ac.uk/user_objects/}} \citep{evans2020}. Our source, identified as LSXPSJ143025.6+415959 in \swift/XRT data, was observed between 2008-08-24 and 2024-05-09 in 68 datasets. Two \swift/XRT spectra are constructed from the available datasets, after splitting them into  two consecutive epochs (see Section \ref{SpectralAnalysis} for details). This grouping is empirically guided by identifying time intervals in which variations in the spectral shape occur, while also ensuring sufficient photon statistics for reliable spectral analysis. 

\section{X-ray spectral analysis}\label{SpectralAnalysis}

\subsection{\xmm\ spectra}\label{individualfits}

Fig. \ref{xmm_spec} presents the PN spectra of 2MASS\,J14302580+4159572 for the three distinct \xmm\ observations of Table \ref{xrayobservations_log}. 
For clarity, the MOS spectra are not shown in this figure, although they are included in the spectral fitting process. The source demonstrates a pronounced spectral change in the 2005 observation, characterised by a strong reduction (factor of about 4) in X-ray flux at energies below $\sim$3\,keV. This variation is already evident in Fig.\,\ref{fig:lc_master} plotting the flux density at 2\,keV as a function of observation epoch. 

We begin the spectral analysis using {\sc xspec} to fit the spectra with a simple power-law ({\sc pl}) model. Both the photon index ($\Gamma$) and the normalisation parameters are tied across the EPIC instruments. Each of the three {\xmm} observation epochs is fitted independently. The analysis yields a typical photon index of $\sim$1.8 \citep[e.g.][]{nandra1994} for both the 2002 and 2003 observations, while the photon index in the 2005 observation drops below 1 (see Table \ref{pl_results}). The reduced $\chi^2$ values are close to unity, which indicates that the model describes well the data within the expected statistical fluctuations.  The corresponding null‑hypothesis probabilities  for the 2002, 2003, and 2005 epochs are 0.5, 0.7, and 0.13 respectively.

Although this modelling is simplistic, it is presented for two reasons. First, the significant discrepancy in photon index values -- identified during an automated fitting process of an extended QSO catalogue -- led to the discovery of this event. Secondly, even this simple model provides important insights into the flux variations associated with  variable or transient phenomena. Indeed, the analysis suggest that the 0.5--3\,keV X-ray flux drops by a factor of four in the 2005 observation, while it remains constant between the other two in 2002 and 2003. Moreover, the flux in the harder part of the spectrum (3–10 keV) remains nearly constant across the three {\xmm} observations. 
Table \ref{pl_results} summarises the single power-law spectral fitting results.

Motivated by the flat spectral shape of the 2005 \xmm\ dataset, we further investigate whether the addition of a neutral absorption component can account for the turnover of the spectrum at energies below $3$\,keV. To this end, we fit only the 2005 dataset using a power-law model modified by a photoelectric absorption component ({\sc zphabs} model in {\sc xspec}). The addition of the neutral absorption component does not change the fitting results. The best fit photon index remains less than one, ($\Gamma<1$), while the estimated hydrogen column density is negligible. If the photon index is fixed to $\Gamma=1.8$, the fit is strongly rejected with a null hypothesis probability of 10$^{-6}$. 

Based on the evidence above, we test a more complex model to interpret the flat slope of the 2005 \xmm\ spectrum, i.e. a partially covering absorber composed of ionised material.  This interpretation is consistent with the presence of WAs, which are generally understood as ionised outflows obscuring the source and producing characteristic soft X‑ray absorption features. The residuals at lower energies (below 0.5 keV) in the 2005 \xmm\ observation (green data in Fig. \ref{xmm_spec}) support this scenario, since ionised gas becomes increasingly transparent at very soft X‑ray energies as the photoelectric opacity decreases with ionisation. 

We model the 2005 \xmm\ spectrum using a power-law modified by absorption from partially ionised material described  by the {\sc zxipcf}  model of {\sc xspec} (i.e. {\sc zxipcf*pl}). {\sc zxipcf} uses a grid of XSTAR \citep{kallman2001} photoionised absorption templates assuming that the absorbing medium covers a fraction of the source. While the covering fraction is a free parameter in the model, we fix it to unity in order to simplify the fitting procedure, thus effectively assuming that the absorber fully covers the source. In Fig. \ref{spec_2005} we present the PN and MOS spectra along with the best-fit model for the 2005 observation, where the significant spectral change is observed. The black line corresponds to the PN spectrum, and the red and green lines to the MOS1 and MOS2 spectra respectively.  The spectral fitting results listed in Table \ref{zxipcf_results} indicate that this model provides an excellent description of the X-ray data. This indicates that the 2005 obscuration event can be attributed to moderately dense ($N_H \approx \rm 4 \times 10^{22}\,cm^{-2}$) and mildly ionised ($\rm \log\,\xi \approx 1.8~\mathrm{erg\,cm\,s^{-1}}$) gas along the line of sight. 
For completeness we also investigate whether the 2002 and 2003 observations are affected by highly ionised material that may not be easily identified due to the low quality of the spectra. Fitting the same model to both the 2002 and 2003 observations we obtain 90\% upper limits for the column density of 1.3 and $1.1 \times10^{21}$ $\rm cm^{-2}$ respectively. For the ionisation parameter  ($\log\,\xi$) we find upper limits of $-0.6$ and $-1$ respectively. These results suggest negligible ionised material compared to the 2005 observation.

\subsection{Joint analysis of {\xmm}, {\nustar}, and {\swift}/XRT spectra}\label{sec:join-analysis}
\
In this section we include in the analysis the two available \nustar\ observations from 2014 and 2023 and the \swift/XRT datasets. Although the spectral change discussed above manifests at energies below the \nustar\ band-pass, these higher-energy data provide nonetheless tighter constraints on the photon index and hence, the column density. The \nustar\  observations also provide independent measurements of the hard-band (3--10\,keV)  X-ray flux across a broader range of epochs. The latter is important to test our assertion, based on \xmm\ data only, that the source’s hard X-ray flux does not change significantly with time. Moreover, \swift/XRT observations offer complementary information on  the observed spectral change, particularly its temporal evolution, by sampling additional epochs not probed by \xmm\.. 
For the \swift/XRT dataset two stacked spectra were constructed after grouping all available observations into two consecutive periods, from  2008 to 2022  and 2023 to 2024. These intervals were selected empirically by stacking consecutive observations in yearly steps until each extracted spectrum contained at least 50 counts, which we adopt as the minimum requirement for reliable spectral fitting. Among the different combinations that satisfied the minimum counts requirement we retained the one that showed the largest change in column density.

Fig. \ref{spec_all} shows all the available X-ray spectra of 2MASS\,J14302580+4159572, fit with a single power-law model, absorbed by partially ionised material, using the {\sc zxipcf}  model in {\sc xspec}. For clarity the MOS data are not displayed, although they were included in the fit. We restrict the analysis to energies below 24\,keV, and ignore higher energies where the \nustar\ effective area declines steeply and the background level increases. Likewise, we omit energies  below 0.3\,keV from the \xmm\ and \swift/XRT spectra as this regime is increasingly affected by  instrumental background and low-energy particle interactions. A Galactic hydrogen column density of $N_{H,\, \rm Gal.}=10^{21} \rm cm^{-2}$ is adopted \citep{HI4PI} using the {\sc tbabs} model. Therefore the complete model in {\sc xspec} terminology is {\sc tbabs*zxipcf*pl}. For each observation, the spectral parameters are tied across the instruments of the same mission (EPIC‑PN and MOS for \xmm\, and FPMA and FPMB for \nustar). The power-law normalisation and the column density are allowed to vary  across epochs.  The photon index, the ionisation parameter (which quantifies how many ionising photons each gas particle receives), and the covering fraction are free to vary during the fitting process but are tied across all observations. This approach minimises overfitting  of the modest-quality X-ray spectra. This setup gives an excellent fit to the data ($\chi^2/dof=222.3/270$) with a corresponding null hypothesis probability of 0.56. The photon index ($\Gamma$) is  $1.92^{+0.09}_{-0.07}$, the ionisation parameter ($\log\xi$) is $1.38^{+0.26}_{-0.24}$ and the covering factor (cf) is $0.92^{+0.08}_{-0.06}$.  All quoted uncertainties represent the 90\% confidence interval. In Table~\ref{final_fits}, we summarise the best fit parameters derived from the simultaneous spectral fitting results of all available observations.

In Fig.  \ref{norm} we explore the intrinsic (i.e. corrected for line-of-sight obscuration) long-term hard-band (3--10\,keV) flux variability of the source over a $\sim$22-year period. For the two \swift/XRT observations, the uncertainties along the time axis represent the temporal span of the observations included in the spectral extraction, covering the periods 2008–2022 and 2023–2024, respectively. The source exhibits no significant evidence of hard X-ray flux variability over the observed period, thereby confirming our initial estimates based on the \xmm\ data (see Fig. \ref{xmm_spec} and Table \ref{pl_results}). In contrast to the relatively stable 3-10\,keV X-ray flux, the simultaneous spectral analysis reveals that the ionised column density exhibits significant evolution in the same 22-year period. In Fig.~\ref{nhvar}, we plot the best-fit ionised column density values derived from the spectral fitting, as a function of observation time. During the first two \xmm\ observations, there is no compelling evidence for ionised material along the line of sight. This is consistent with the individual spectral fitting results presented in Section \ref{individualfits}.  However, in the third (2005) \xmm\ observation, the source appears obscured by an ionised column density of approximately $5\times10^{22}~\rm cm^{-2}$. This obscured state remains largely unchanged, within the uncertainties, in the first \swift/XRT spectrum covering 2008–2022. The second \swift/XRT spectrum (2023–2024) suggests a decrease in the column density, although the uncertainties of the two \swift/XRT spectra alone do not permit a strong conclusion ($\sim2\sigma$ confidence level).  The comparison of the obscuration seen in the 2005 \xmm\ spectra and the 2023–2024 \swift/XRT spectrum favours a decrease in the obscuration at a 3$\sigma$ confidence level.   We note that, without additional temporal coverage we cannot firmly determine whether the source has fully returned to an unobscured state.  

The \nustar\ spectra are essentially unaffected by the absorption feature, owing to the instrument’s energy band-pass, which is limited to energies above 3\,keV. Indeed, the \nustar\ data provide only upper-limits to the column density. The absence of any detectable absorption signature in the \nustar\ band  indicates that the true column density of the source during the \nustar\ observation epoch is unlikely to be significantly  higher than the maximum measured value.

\begin{table}[ht]
\caption{Fitting results of the {\sc pl} model in the three \xmm\ observations.}
\label{pl_results}
\centering
\begin{tabular}{lcccc}
\hline
 Year   & $\Gamma$ & $ F_{\rm 0.5-3\,keV}$  & $F_{\rm 3-10\,keV}$   & $\chi^2$/dof \\
  &  &  ($\rm erg~s^{-1}~cm^{-2}$) & ($\rm erg~s^{-1}~ cm^{-2}$) &  \\

\hline
2002 & 1.79$^{+0.11}_{-0.11}$ & $1.50\times10^{-13}$  & $1.39\times10^{-13}$ & 97.53/90 \\ 
2003 & 1.85$^{+0.06}_{-0.06}$ & $1.48\times10^{-13}$  & $1.24\times10^{-13}$ & 66.69/60 \\
2005 & 0.95$^{+0.12}_{-0.12}$ & $4.37\times10^{-14}$  & $1.33\times10^{-13}$ & 45.66/36 \\
\hline
\end{tabular}
\tablefoot{We list the inferred photon index ($\Gamma$), the observed fluxes in the 0.5--3\,keV and 3--10\,keV bands and the $\chi^2$/dof of the best-fit.}

\end{table}

\begin{figure}[ht]
\centering
\includegraphics[width=1.20\columnwidth]{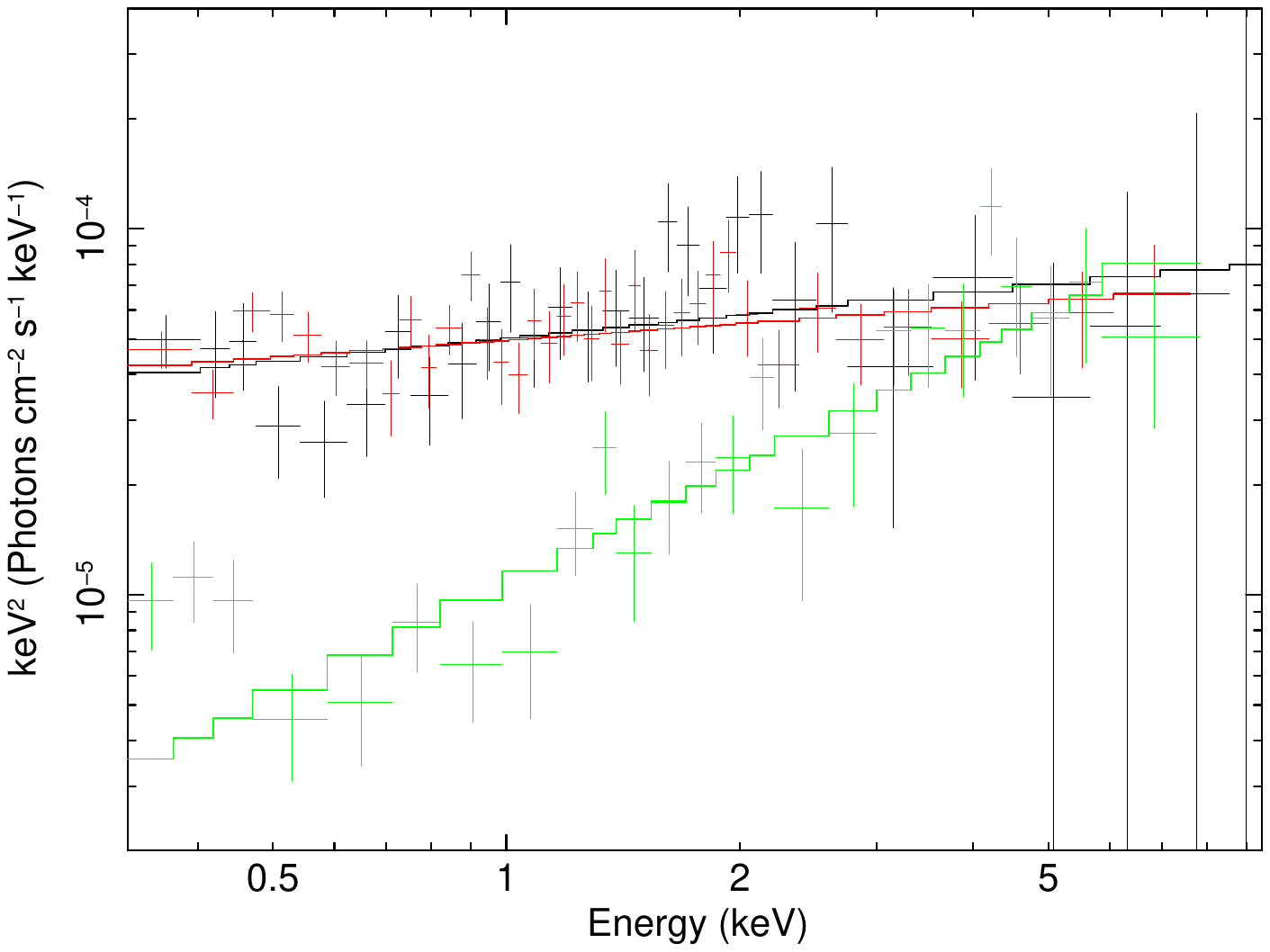}
\caption{EPIC PN spectra of 2MASS\,J14302580+4159572. The black data correspond to the 2002 observation, the red to the 2003 observation, and the green to the 2005 observation. The MOS spectra are not shown for clarity  although they were included in the fitting. The solid lines show the best fit {\sc pl} model for each epoch.}
\label{xmm_spec}
\end{figure}

\begin{figure}
\begin{center}
\includegraphics[width=1.20\columnwidth]{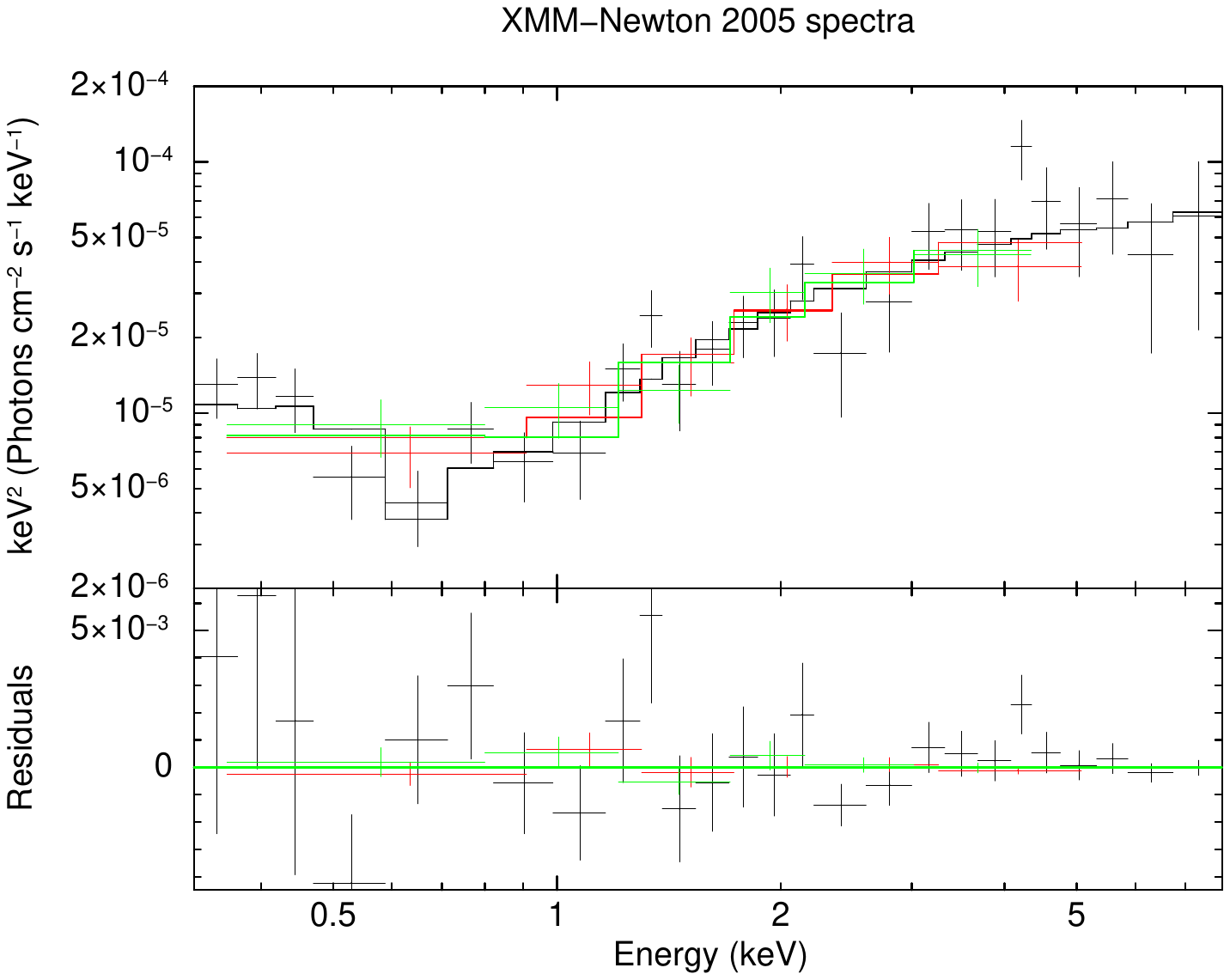}
\end{center}
\caption{EPIC PN and MOS spectra of 2MASS\,J14302580+4159572 for the 2005 observation. The black line corresponds to the PN spectrum, and the red and green lines to the MOS1 and MOS2 spectra, respectively. The solid curves show the best-fit {\sc zxipcf*pl} model.}
\label{spec_2005}
\end{figure}

\begin{table*}[ht]
\caption{Fitting results to the 2005 \xmm\ observation using the {\sc zxipcf}*{\sc pl} spectral model.}
\label{zxipcf_results}
\centering
\begin{tabular}{ccccccc}
\hline
$\rm N_H$ & $\rm \Gamma$ & $\log(\,\xi)$ & cf & $F_{\rm 0.5-3\,keV}$ & $F_{\rm 3-10\,keV}$ & $\chi^2$/dof \\
$\times 10^{22}~\rm cm^{-2}$ & & & & $\rm ergs~s^{-1}~cm^{-2}$ & $\rm ergs~s^{-1}~cm^{-2}$ & \\
\hline
$4.1^{+2.28}_{-1.78}$ & 1.60$^{+0.38}_{-0.23}$ & $1.83^{+0.25}_{-0.44}$ & 1 & $4.50\times10^{-14}$ & $1.08\times10^{-13}$ & 28.81/38 \\
\hline
\end{tabular}
\tablefoot{We list the inferred column density ($N_H$), photon index ($\Gamma$), ionisation parameter ($\log\,\xi$), and  $\chi^2$/dof of the best-fit. The covering factor is fixed to unity. The observed fluxes in the 0.5--3\,keV and 3--10\,keV bands are also shown.}

\end{table*}

\begin{figure}
\begin{center}
\includegraphics[width=1.15\columnwidth]{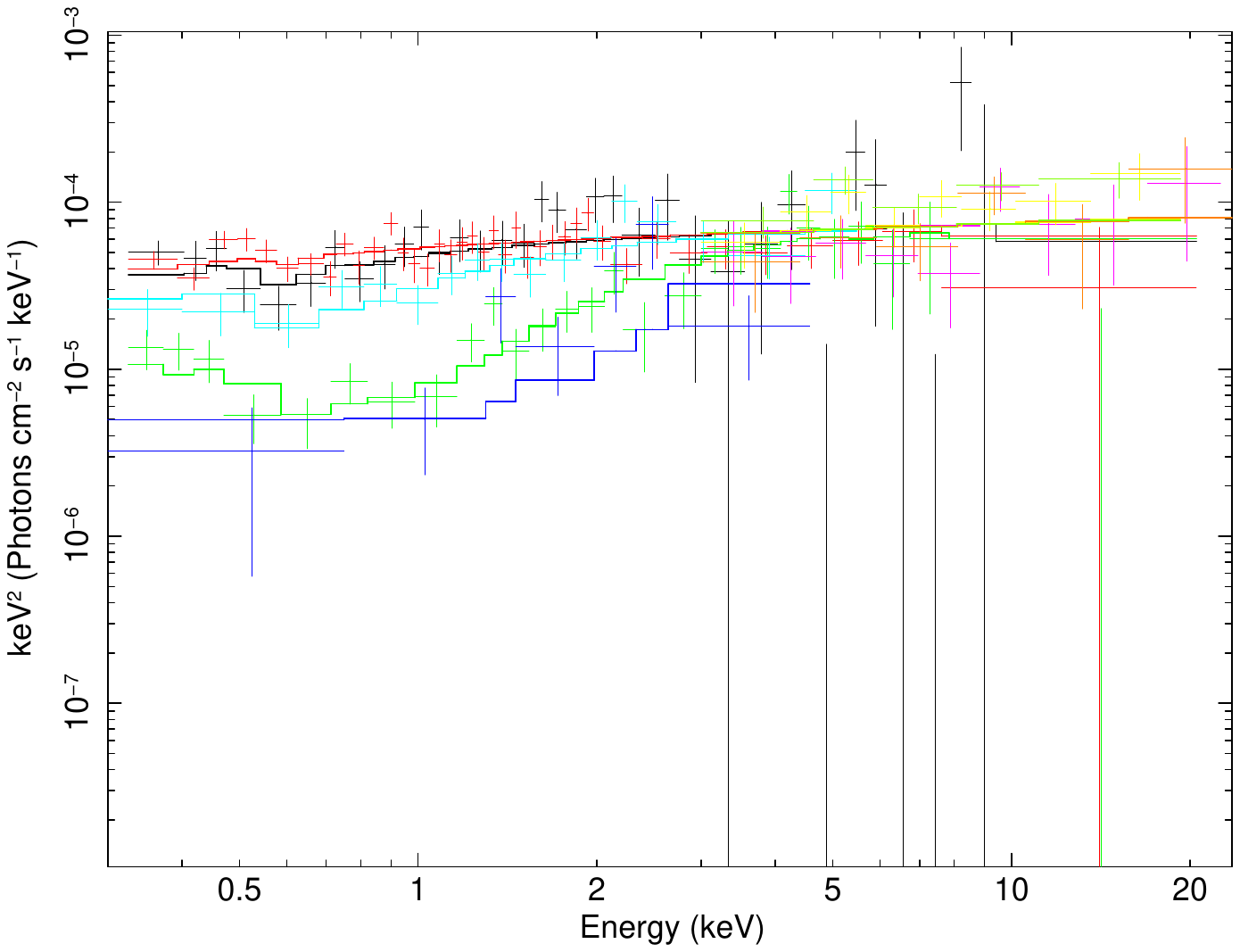}
\end{center}
\caption{Complete set of X-ray spectra for 2MASS\,J14302580+4159572.  
The black, red, and green data correspond to the 2002, 2003, and 2005 \xmm\ observations, respectively. The blue and cyan spectra correspond to the \swift/XRT observations from the 2008–2022 and 2023–2024 epochs. The solid curves show the best-fit {\sc tbabs*zxipcf*pl} model.}
\label{spec_all}
\end{figure}

\begin{figure}
\begin{center}
\includegraphics[width=1.05\columnwidth]{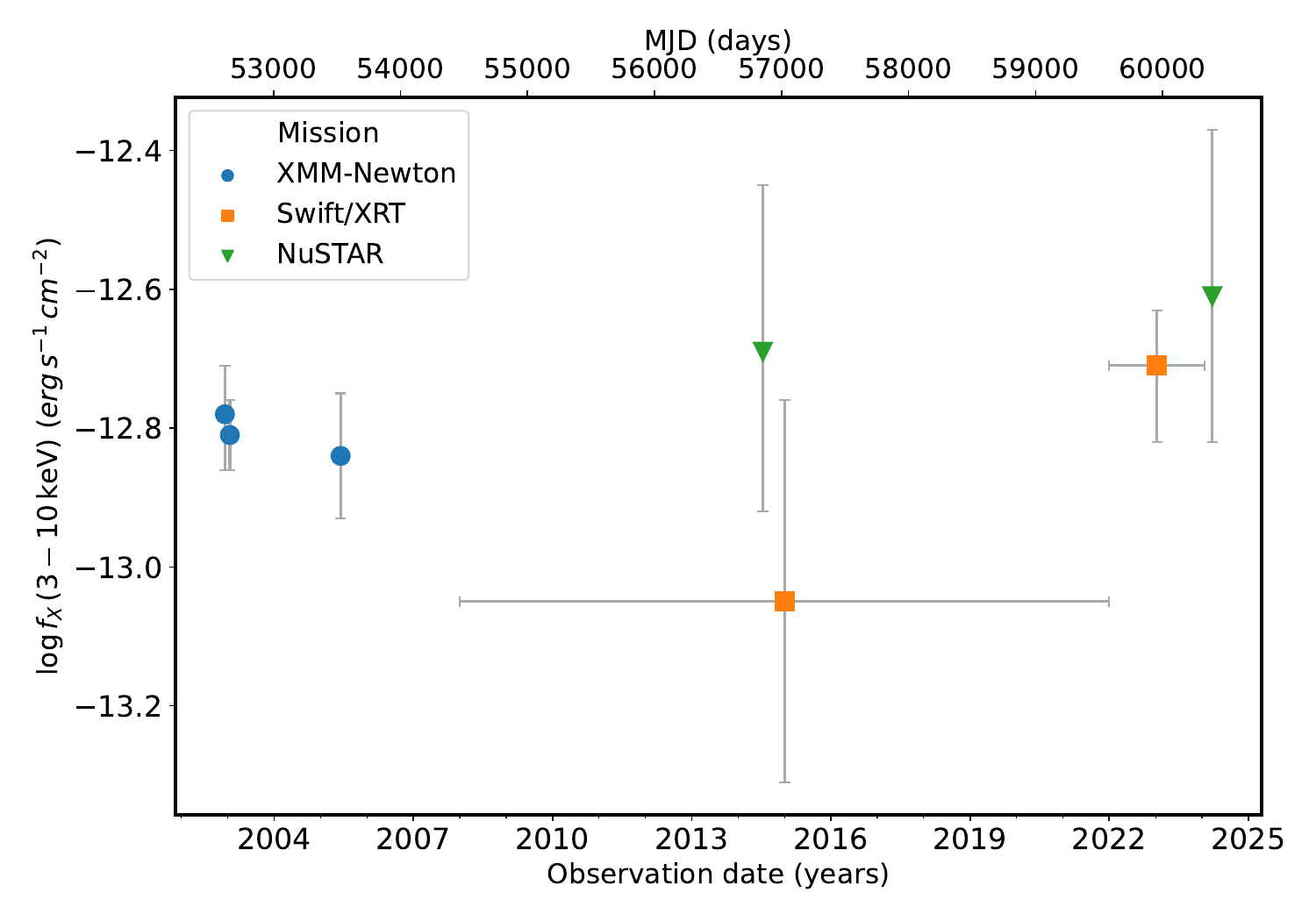}
\end{center}
\caption{Intrinsic 3–10\,keV flux estimates over the 22-year monitoring period. The flux values are derived from the simultaneous spectral-fitting analysis and are shown as a function of observation time. Overall, the flux remains consistent with being unchanged within the error bars across the $\sim$22-year baseline.}
\label{norm}
\end{figure}

\begin{figure}
\begin{center}
\includegraphics[width=1.0\columnwidth]{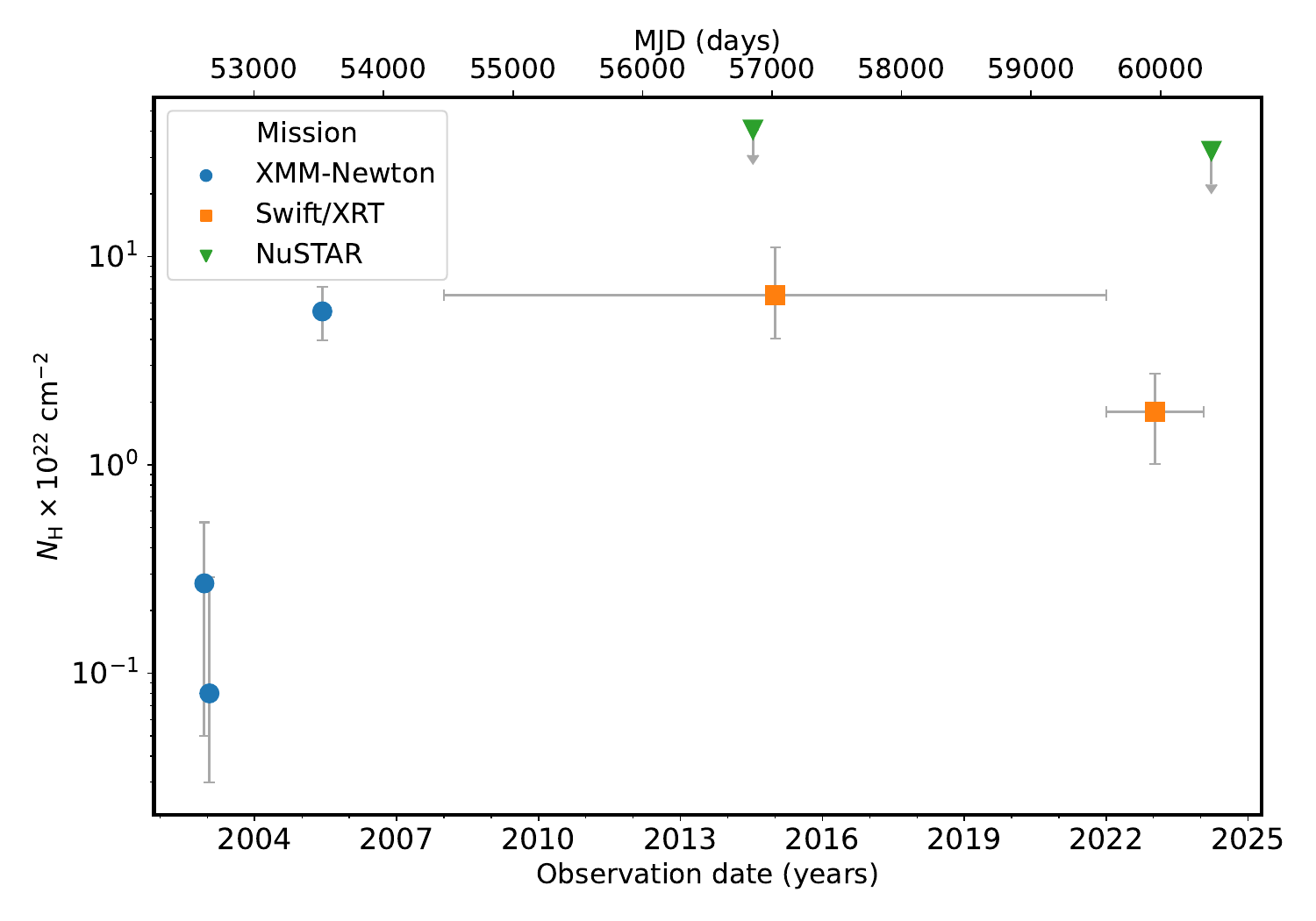}
\end{center}
\caption{Ionised column density as a function of observation epoch.  
The values are inferred from the joint X-ray spectral analysis of Section~\ref{sec:join-analysis}. The source shows significant variability in ionised absorption: during the first two \xmm\ observations the obscuration is low, while in the third \xmm\ epoch the ionised absorption increases. The enhanced obscuration persists through the first \swift/XRT spectrum (2008–2022) but appears to decrease in the stacked 2023–2024 \swift/XRT observations. The \nustar\ estimates are also shown as upper limits for completeness.}
\label{nhvar}
\end{figure}

\begin{table*}[ht]
\caption{Joint fitting results of all the available spectra using the {\sc tbabs}*{\sc zxipcf}*{\sc pl} model.}
\label{final_fits}
\centering
\begin{tabular}{lccccccc}
\hline
Mission & Date &  $\rm N_H$  & $\rm \Gamma$ & $log\,\xi$  & cf & logF$_{3-10~\rm keV}$  & $\chi^2$/dof \\
       &       &  $\times10^{22}\,\rm cm^{-2}$ &&   &   $\mathrm{erg\,cm\,s^{-1}}$ &  $\rm ergs~s^{-1}~cm^{-2}$ &    \\
\hline
\multirow{3}{*}{\xmm} & 2002-12-09 & $0.27^{+0.26}_{-0.22}$ & \multirow{7}{*}{$1.92^{+0.09}_{-0.07}$} & \multirow{7}{*}{$1.38^{+0.26}_{-0.24}$} & \multirow{7}{*}{$0.92^{+0.08}_{-0.06}$} & $-12.90^{+0.07}_{-0.08}$ & \multirow{7}{*}{222.3/270} \\
                    & 2003-01-17  & $0.08^{+0.21}_{-0.05}$ &    &  &   & $-12.93^{+0.05}_{-0.05}$ &     \\
                    & 2005-06-05  & $5.46^{+1.71}_{-1.50}$ &    &  &   & $-12.96^{+0.08}_{-0.10}$ &     \\
\multirow{2}{*}{\swift/XRT} & 2008-2022 & $6.54^{+4.56}_{-2.58}$ &&  && $-13.10^{+0.29}_{-0.24}$ &     \\
                           & 2022-2024 & $1.80^{+0.95}_{-0.79}$ &&  && $-12.84^{+0.08}_{-0.10}$ &     \\
\multirow{2}{*}{\nustar}    & 2014-07-14 & (40.60) &&  && $-12.80^{+0.19}_{-0.27}$ &     \\
                           & 2023-12-09 & (32.27) &&  && $-12.75^{+0.20}_{-0.23}$ &     \\
\hline
\end{tabular}
\tablefoot{We list the inferred column density ($N_H$), the photon index ($\Gamma$), the ionisation parameter ($\log\,\xi$), the covering factor (cf) and the  $\chi^2$/dof of the best-fit. The intrinsic fluxes in the 3--10\,keV band are also shown. The photon index ($\Gamma$), the ionisation parameter ($\log\,\xi$), and the covering fraction (cf) are kept fixed across all observations. For the \nustar\ observations, the $N_H$ value corresponds to the 90 per cent  upper limit.}
\end{table*}

\section{Optical photometry}\label{opticalanalysis}

We explore the level of optical variability by estimating the structure function \citep[SF;][]{Kozlowski2016} of the ZTF $g$ band light curve of Fig. \ref{fig:lc_ztf}. Fig. \ref{fig:sf_ztf} compares the SF\footnote{We adopt the definition ${\rm SF}(\Delta t)=\sqrt{\langle(m_i-m_j)^2\rangle-2\langle \sigma^2\rangle}$,  where $m_i$ and $m_j$ are the optical magnitudes at epochs $i$, $j$ separated by the rest-frame time interval $\Delta t$ and $\sigma^2$ represents the photometric uncertainty at each epoch. Different definitions in the literature give similar results.} measurements at different rest-frame timescales with the ensemble of the SDSS QSO population \citep{MacLeod2012}. The observed optical flux variations of 2MASS\,J14302580+4159572 are consistent within the uncertainties with the expectation for QSOs. In contrast to X-rays, the source does not appear to be exceptionally variable at optical wavelengths, at least not during the ZTF epochs. Comparison of the ZTF photometry with the historical SDSS-DR7 $g$ band magnitude (see Fig. \ref{fig:lc_ztf}) suggests an average offset of about 0.2-0.3\,mag, which is consistent with the structure function expectation.  

We also fit templates to the observed optical/infrared Spectral Energy Distribution (SED) of 2MASS\,J14302580+4159572 to explore the presence of dust that may be associated with the ionised absorber at X-rays. As explained in Appendix \ref{sec:sed} we compile two sets of optical photometric observations. The first corresponds to the X-ray unobscured phase and uses the SDSS-DR7 and OM observations (separated by a few months).  The second uses non-simultaneous photometric data from GALEX and LS10 taken in the period between 2005 and 2022, when the source is in the X-ray obscured phase. These two sets are supplemented by the same near-infrared data 
(2MASS\footnote{Two Micron All Sky Survey; \citealp{Skrutskie2006}} and 
UKIDSS\footnote{UKIRT Infrared Deep Sky Survey; \citealp{Lawrence2007}}) 
and mid-infrared data \citep[WISE\footnote{Wide-field Infrared Survey Explorer};][]{Wright2010}, 
which provide information on the host-galaxy contribution and the torus emission, 
under the assumption that these longer wavelengths are not affected by the variations in X-ray absorption.
 We estimate  $E(B-V)=0.15\pm0.02$ for the "unobscured phase" and  $E(B-V)=0.20\pm0.02$ for the "obscured" period. Given the estimated $1\sigma$ uncertainties and the SED construction assumptions, there is negligible temporal variation in the dust extinction to the central engine. The appearance of the X-ray ionised absorption does not have a strong impact on the optical part of the broad-band SED of 2MASS\,J14302580+4159572.

\begin{figure}
	\includegraphics[width=0.9\columnwidth]{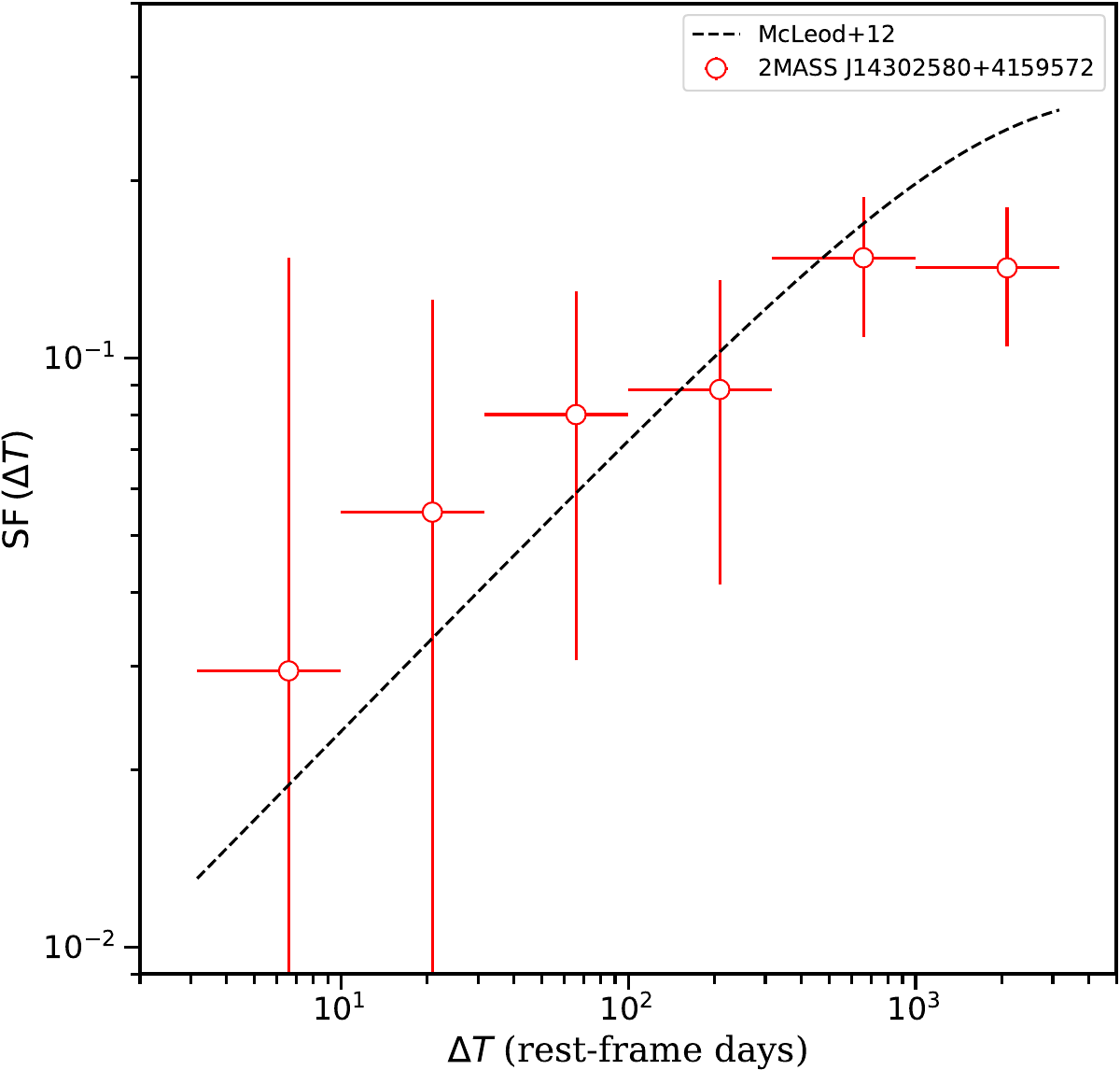}
    \caption{Structure function of the ZTF $g$ band light curve of 2MASS\,J14302580+4159572. The structure function is shown as a function of rest-frame timescale (red open circles). The dashed line represents the analytic fit to the $g$ band structure function of SDSS QSOs presented by \protect\cite{MacLeod2012}.}

    \label{fig:sf_ztf}
\end{figure}

\section{Optical spectra}

\begin{figure*}
	\includegraphics[width=1\textwidth]{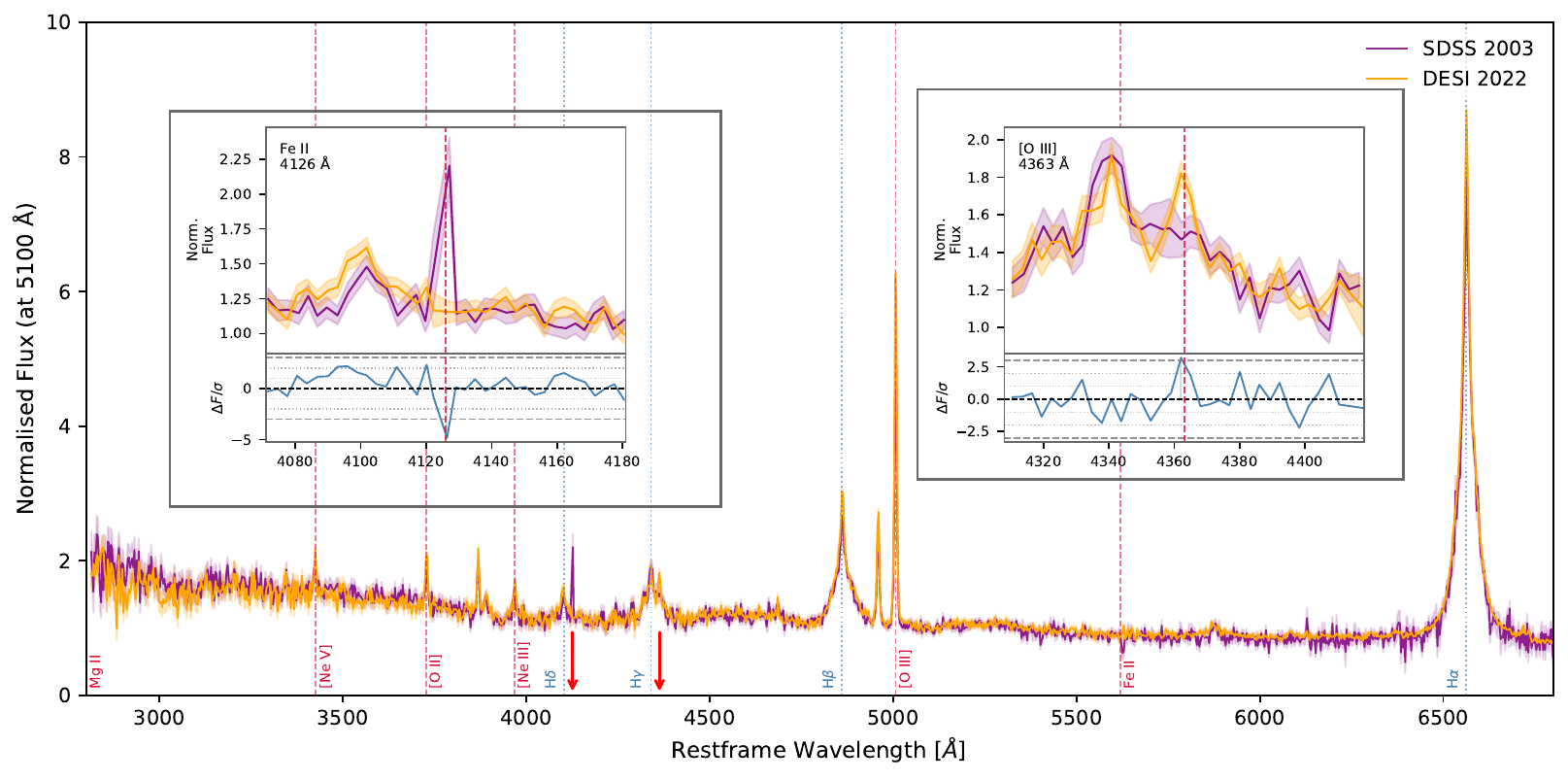}
    \caption{Rest-frame optical spectra of 2MASS\,J14302580+4159572 from SDSS DR7 (purple, 2003) and DESI DR1 (orange, 2022). Both spectra are binned for clarity and normalised at 5100\,\AA. Shaded regions indicate the $1\sigma$ photometric uncertainties. Major AGN emission lines are indicated with dashed crimson lines, including high-ionisation forbidden lines ([Ne\,V]\,$\lambda$3426\,\AA, [O\,III]\,$\lambda$4363\,\AA, [O\,III]\,$\lambda$5007\,\AA), low-ionisation Fe\,II complexes ($\lambda$4126--5620\,\AA), and Mg\,II\,$\lambda$2800\,\AA. Balmer lines (H$\delta$, H$\gamma$, H$\beta$, H$\alpha$) are marked with dotted blue lines. Arrows indicate the [O\,III]\,$\lambda$4363\,\AA\ and Fe\,II\,$\lambda$4126\,\AA\ lines that show significant changes  ($>3\sigma$) between the two epochs. The inset plots zoom into the spectral regions of the two lines. The lower panels of each inset plot show as a function of wavelength the flux difference significance ($\Delta F/\sigma$) between the two epochs.}
    \label{fig:optical_specta}
\end{figure*}

In Fig.~\ref{fig:optical_specta}, we present the optical spectra of 2MASS\,J14302580+4159572 obtained with SDSS (unobscured X-ray epoch) and DESI (obscured X-ray epoch). Both spectra are rebinned to 3\,\AA\ resolution and are normalised at 5100\,\AA. We then estimate the difference between the two spectra (DESI-SDSS) to explore changes in optical spectral features. We find significant (i.e. $>3\sigma$) changes in the high-ionisation [O\,III]~$\lambda4363$ line, as well as in the low-ionisation Fe\,II feature around 4126\,\AA. These changes are shown in detail in the insets of Fig.~\ref{fig:optical_specta}. The lower panels of each inset plot show the flux-difference significance ($\Delta\,F/\sigma$) between the two epochs for each spectral element bin. This is computed as $(F_{DESI} - F_{SDSS})/(\sigma_{DESI}^2 +\sigma_{SDSS}^2)^{0.5}$. 

The comparison reveals an enhancement of the  [O\,III]~$\lambda4363$ line in the DESI observation. This line originates in compact, highly ionised gas located outside the BLR and extending into the inner NLR region \citep[e.g.][]{Nagao2001,Baskin2005}. The appearance and strengthening of the line in the DESI spectrum suggest a connection to the ionised absorber visible at X-rays. 

Additionally, the SDSS spectrum shows a strong, low-ionisation, Fe\,II emission feature near 4126\,\AA\ \citep{Sarkar2021}, which is strongly suppressed in the DESI epoch. The Fe\,II emission near 4126\,\AA\ originates predominantly in the broad-line region (BLR), where the combination of high densities and low ionisation parameters allows Fe\,II to survive. As the ionisation parameter increases, Fe\,II is efficiently destroyed, leading to the suppression of Fe\,II emission observed in the DESI  epoch \citep[e.g.][]{Gaskell2022}. However, we note that the Fe\, II emission features could arise from numerous transitions of the complex Fe\,II ion, whose detailed origin and excitation pathways remain only partially understood. Their presence and strength depend sensitively on the ionisation level, microturbulence effects,  gas density, and the shape of the spectral energy distribution \citep{kovacevits2010}. We therefore offer only a tentative interpretation, guided by the observed X‑ray behaviour. }

\section{Modelling}

Changes in AGN obscuration are attributed to  several physical mechanisms. These include intrinsic column-density variability caused by clouds moving into or out of the line of sight {\citep[e.g.][]{risaliti2002, lamer2003}}, variations in the ionisation state of the obscuring material in response to changes of the AGN’s luminosity { \citep[e.g.][]{risaliti2005}}, and AGN-driven outflows manifested as WAs {\citep[e.g.][]{mehdipour2017, kara2021}}. Here, we briefly revisit these scenarios in the context of our observational findings for 2MASS\,J14302580+4159572 and offer some insights into their possible relevance, while acknowledging the limitations of our data.

\subsection{Changes in the AGN luminosity}

According to this scenario an increase of the AGN luminosity could lead to an enhancement in the ionisation state of the obscuring material, making it more transparent to X-ray radiation, which would in turn lead to an apparent decrease of the line-of-sight column density. However, this process requires substantial changes in the intrinsic luminosity 
 \citep[e.g.][]{risaliti2005} and may not apply to  most cases seen in AGN, where the changes in the obscuration are not typically correlated with large variations of the bolometric luminosity (but see \citealp{mehdipour2022}).  In the case of 2MASS\,J14302580+4159572,  both the X-ray continuum (see Fig. \ref{norm}) and the optical flux (see Figs. \ref{fig:lc_ztf}, \ref{fig:lc_master}, \ref{fig:sf_ztf}) do not show strong variation with time and therefore the changing luminosity scenario is disfavoured. 

\subsection{Cloud scenario}\label{sec:orbiting-cloud}

An alternative plausible scenario to explain the observed changes is intrinsic column‑density variability, caused by clouds moving into and out of the line of sight. While occultations of the X-ray source by intervening clouds are mainly observed in obscured AGNs \citep[e.g.][]{risaliti2002, akylas2002,Liu2021}, examples of this phenomenon are also reported for type~1 sources. \citet{risaliti2011} present strong spectral variability in the narrow-line Seyfert-1 Mrk\,766 during a 7-day \xmm\, monitoring campaign. They show that these changes can be interpreted as eclipses by BLR clouds crossing the line of sight to the X-ray source. Also \cite{lamer2003} attribute the unusual hardening of the 2-10\,keV X-ray spectrum of NGC\,3227, which lasted several months, to motions of clouds close to the BLR region. Furthermore, \citet{matt2011} analyse the X-ray spectra of the Seyfert 1 galaxy Mrk 704, observed twice by \xmm\, three years apart. They report significant changes in the properties of the absorbing material which however, could only partly be attributed to the presence of orbiting clouds around the nucleus. 

We consider an idealised scenario where such an obscuring medium is composed of individual, consecutive, non-overlapping  clouds of radius $R_{\rm cloud}$, with constant ionisation state and uniform density \citep[e.g.][]{lamer2003, lian2025}. Although these assumptions are certainly approximations, they allow order-of-magnitude  estimation of the cloud properties, particularly when their sizes are much smaller compared to their distances from the illuminating source.  

A cloud of gas orbiting a non-rotating (Schwarzschild) black hole with mass $M_{BH}$ at radius $R$, has an orbital velocity  ($v_{cloud}$) approximately given by

\begin{equation}
  v_{cloud} = \sqrt{\frac{G\cdot M_{BH}}{R}}. 
        \label{eq:vorb1}
\end{equation}

\noindent We further assume that the cloud moves perpendicularly into the line of sight, progressively increasing the obscuration of the central source. 
The obscuration rises from nearly zero in the 2003 \xmm\ observation to its highest value in the 2005 \xmm\ observation over a time interval defined as $t_{min}$ . We interpret this interval  as marking the moments when the cloud first enters the line of sight and when the line of sight crosses the cloud's centre, where the column density attains its maximum value.
In this case the cloud velocity is related to its radius, $R_{cloud}$, through the relation

\begin{equation}
         v_{cloud} = \frac{R_{cloud}}{t_{min}}.
        \label{eq:vorb2}
\end{equation}

\noindent Moreover, the diameter of the cloud ($D_{cloud}=2\cdot R_{cloud}$) can be approximated by $N_H/n$, where $N_H$ is the observed column density of the cloud and $n$ is its density. Therefore the velocity of the cloud across the line of sight can be expressed as 

\begin{equation}
v_{cloud} = \frac{D_{cloud}}{2\cdot t_{min}}=\frac{N_H}{2\cdot n \cdot t_{min}}.
        \label{eq:vorb3}
\end{equation}

\noindent From the definition of the ionisation parameter we have  

\begin{equation}
    \xi=\frac{L_{ion}}{n\cdot R^2},
    \label{eq:xi}
\end{equation}

\noindent where $L_{ion}$ is the ionising luminosity calculated between 13.6\,eV and 13.6\,keV and $R$ is the distance between the cloud and the ionising source. Substituting the density of the cloud from Equation \ref{eq:vorb3} and the cloud velocity from Equation \ref{eq:vorb1} into Equation \ref{eq:xi} and solving for $R$, we find 

\begin{equation}
R = (G\cdot M_{BH})^{1/5}\frac{({2\cdot t_{min}\cdot L_{ion})}^{2/5}}{{(N_H\cdot \xi)}^{2/5}}.
        \label{eq:R_cloud}
\end{equation}

\noindent Using Equation \ref{eq:R_cloud}, we can estimate the distance $R$ of the cloud from the  central black hole. We use a black-hole mass value of $\log (M_{\rm BH}/M_\odot)=8.35\pm0.5$ based on the SDSS optical spectrum and the methods described in \cite{wu2022}. Using the time span between the low and high obscuration epochs in  Fig. \ref{nhvar} we adopt $t_{min}\simeq2.5\rm \,yr$. The values of the ionisation parameter and the maximum column density are listed in Table \ref{spec_all}. Finally, we estimate $L_{ion}=2\times10^{44}\rm \, erg\,s^{-1}$ by integrating the best fit X-ray spectral model in the interval 0.0136-13.6\,keV. 

Substituting these values to Equation \ref{eq:R_cloud} we find $R\simeq1.7^{+1.5}_{-0.8}\times10^{18} \rm \, cm$ ($\simeq1.8$\,light-years or $\simeq0.55$\,pc). At this distance, Equation \ref{eq:xi} yields a cloud density of $n\simeq3\times10^6\rm \,cm^{-3}$. From Equation \ref{eq:vorb3} we estimate the cloud velocity $v_{cloud}\simeq1321^{+412}_{-256}\rm \, \, km\,\,s^{-1}$ and Equation \ref{eq:vorb2} gives a cloud radius of about 4 light-days.  If we further assume, based on Fig. \ref{nhvar}, that the maximum duration of the eclipse is roughly 20\,years, then an upper limit for the size of the obscuring structure is approximately between 30-40 light-days.  These results indicate that the putative clouds are most probably located outside the BLR.

\subsection{Outflow scenario}

Unobscured (Type~1) AGNs can exhibit strong spectral effects from variable, ionised gas along the line of sight. Such ionised absorbers are commonly associated with AGN outflows and are  identified as absorption troughs in both X‑ray and optical/UV spectra. These outflows are often launched from the accretion disc and can produce temporary X‑ray obscuration events, as observed in a handful of nearby Seyferts. The obscurer in such events is typically moderately ionised, has column densities of order $\rm 10^{22}~cm^{-2}$ and partially covers the direct X-ray emission from the central source on timescales ranging from months to decades or even longer.  For example, \cite{mehdipour2017} and \cite{kara2021} report large changes in the X‑ray absorption of NGC 3783 and Mrk 817, respectively, on month-timescales. The properties of this obscuration are consistent with winds launched from the accretion disc. Also, \cite{kaastra2014} and \cite{ebrero2016} present evidence for longer scale (years) obscuration events in NGC5548 and NGC985, respectively,  attributed to clumpy stream of ionised gas  from the accretion disc reaching beyond the BLR.  

We explore the possibility that the observed change in the soft X‑ray flux of  2MASS\,J14302580+4159572 is produced by an obscuration event associated with an outflow. As in the cloud scenario, we adopt an idealised model in which the obscuring medium has a constant ionisation state and a uniform density. Although these assumptions are simplifications, they allow  us to quantify the outflow properties.   

The maximum distance the outflow reaches can be estimated if both its velocity and duration are known. From Fig. \ref{nhvar}, which shows the evolution of the column density, we see  a rapid rise lasting about 2–3 years, after which the column density peaks and then stays roughly constant for about 18 years — likely due to continuous replenishment by the wind— before eventually beginning to decline. The rising time of 2-3 years gives a rough estimate on the effective travel time of the outflowing gas i.e. before it becomes too dilute to continue. This is $\sim$2.5 years or $\sim 7.9\times10^7$ sec. On the other hand the outflow velocity remains unconstrained. The limited EPIC spectral quality and the lack of sufficient signal-to-noise ratio  RGS data do not allow the identification of ionised absorption spectral features and the estimation of the outflow velocity. We therefore assume an outflow velocity  equal to the escape velocity, $v_{esc},$ at distance $R$ from the central source, which represents a meaningful lower bound.

\begin{equation}
         v_{esc} = \sqrt{\frac{2\cdot G\cdot M_{BH}}{R}}. 
        \label{eq:v}
\end{equation}

\noindent  We can further estimate the maximum distance, $D$, reached by the outflow using

\begin{equation}
         D = v_{esc}\cdot t.
         \label{eq:R}
\end{equation}

\noindent The maximum column density, $N_H$, can be approximated as

\begin{equation}
    N_H = n \cdot D,
    \label{eq:nh1}
\end{equation}

\noindent where $n$ is the density  of the outflowing material. Using again equation \ref{eq:xi} to  estimate $n$ and  combining Equations \ref{eq:v}, \ref{eq:R} and \ref{eq:nh1} we obtain an approximate expression for the launch radius of the outflow 

\begin{equation}
R = (2\cdot G\cdot M_{BH})^{1/5}\frac{({t_{min}\cdot L_{ion})}^{2/5}}{{(N_H \cdot \xi)}^{2/5}}.
        \label{eq:R_outflow}
\end{equation}

\noindent  This result is similar to Equation \ref{eq:R_cloud}, which gives the distance of a cloud orbiting the source, because the escape velocity adopted here is comparable to the cloud’s Keplerian velocity used in the cloud scenario (see Equation \ref{eq:vorb3}). Using the same values for $M_{BH}$, $t_{min}$, $\xi$, $N_H$ and $L_{ion}$ as in Section \ref{sec:orbiting-cloud}, we estimate $R=1.4^{+1.5}_{-0.6}\times10^{18}\, \rm cm$, ($\simeq1.5$\,light-years or $\simeq0.45$\,pc) outflow (escape) velocity  $2000^{+500}_{-450}\rm \, \,km\,\,s^{-1}$ and $n\approx4\times~10^6\rm~cm^{-3}$. 

\section{Discussion}

In this study, we quantify and characterise the warm absorption variability observed in the type-I QSO 2MASS\,J14302580+4159572. Our data suggest that the absorbing column density initially increases 
from a value consistent with zero within the uncertainties to $N_{\rm H} \approx 6 \times 10^{22}\,\mathrm{cm^{-2}}$  with an ionisation parameter of $\log\,\xi \approx 1.4$ and then declines over a time scale of about 22~years. During this period, the  intrinsic hard (3-10 keV) X-ray luminosity of the source remains constant. Our analysis predicts a characteristic distance of approximately 1-2 light‑years  between the ionised material and the supermassive black hole. In the models we consider, this distance  may correspond either to the trajectory of an obscuring cloud or to the launch site of an outflow event. Such distances effectively place the absorber  beyond the BLR. Indeed, an order-of-magnitude estimate of the size of the BLR using the optical luminosity of 2MASS\,J14302580+4159572 \citep[e.g.][]{wu2022} is $10^{17}\,\rm cm$ or roughly one-light-month \citep[e.g. Equation 2 of][]{kaspi2005}. 

\subsection{The nature of the obscurer}

The presence and long-term persistence of orbiting clouds in an otherwise unobscured type-I AGN present a challenge for the orbiting-cloud interpretation. In this scenario, the obscuring material is unlikely to originate in the dusty torus and is instead more plausibly associated with clumps embedded in outflows \citep[e.g.][]{lian2025}. However, \citet{risaliti2011} show that occultation events caused by clouds crossing the line of sight can occur even in type~1 objects. In their study of the narrow-line Seyfert-1 Mrk\,766, the obscuring clouds --- interpreted as part of the BLR --- exhibit densities of $10^{10}$--$10^{11}\,\mathrm{cm^{-3}}$ and linear dimensions of order $10^{12}$--$10^{13}\,\mathrm{cm}$.
These findings contrast with ours, as the variability we observe unfolds over several years rather than the rapid, hour-scale occultation events reported  in \citet{risaliti2011}.
Recently, an obscuration event was detected in the Seyfert~1 galaxy EC\,04570$-$520 \citep{Markowitz2024}. Using eROSITA to monitor large flux changes in AGNs between consecutive sky scans, the authors found that the soft X-ray flux of EC\,04570$-$520 dipped abruptly for about 10--18 months during 2020--2021, recovered, and then dropped again in early 2022. The 2020--2021 event was caused by a cloud with a column density of $10^{22}\,\rm cm^{-2}$ and a covering fraction of 60\%. The 2022 event involved a cloud with a column density of $3\times10^{23}\,\rm cm^{-2}$ and a covering fraction near 80\%.
Interestingly, the optical/UV continuum flux shows minimal variability. An important difference between the obscuration events in 2MASS\,J14302580+4159572 and EC\,04570$-$520 is that the absorber in the latter is neutral and the optical emission-line spectra do not show changes. Moreover, in our case, the initial 2.5-year period, during which the column density rises to its peak, appears to be followed by a longer phase of roughly 18 years with nearly constant column density. In the context of the orbiting cloud scenario such behaviour would imply a highly elongated clumpy structure with an ellipticity close to 0.9.

In the outflow scenario, the continuous replenishment of the obscuring material can be more naturally explained by the sustained motion of the outflowing gas. In this case, the duration of the constant-obscuration phase does not correspond to the physical size of a single cloud, but instead reflects the timescale over which the outflow remains active and intersects the line of sight. Previous studies frequently report cases in which the X-ray spectrum is significantly affected by variable ionised absorbing gas along the line of sight, commonly associated with AGN outflows and routinely identified through absorption troughs in both the X-ray and optical/UV regimes \citep[e.g.][]{king2015}. Notable examples include NGC\,5547 \citep{kaastra2014}, NGC\,3783 \citep{mehdipour2017}, Mrk\,817 \citep{kara2021}, and NGC\,3516 \citep{mehdipour2022}, where detailed investigations reveal variability on timescales from days to years and place the outflowing gas at distances of a few light-days to light-months from the central source.
If our event is associated with an outflow, then it should be located at a distance of $\sim 1$--$2$ light-years from the black hole, with an estimated outflow velocity of at least several hundred km\,s$^{-1}$. The outflow appears to remain active--that is, continuously replenished or sustained--for roughly 18 years before beginning to decline.

\subsection{Optical spectroscopic information}

A comparison of the SDSS and DESI optical spectra obtained in 2003 and 2022, respectively, before and after the appearance of the X-ray obscuration, reveals subtle differences, such as the appearance of the high-ionisation [O\,III]~$\lambda4363$ line in the DESI spectrum (Fig.\,\ref{fig:optical_specta}). This difference suggests a higher level of ionisation at the transition zone between the BLR and the NLR, where [O\,III]~$\lambda4363$ originates, possibly associated with the emergence of the ionised X-ray absorber. In the scenario of outflowing material for example, we may witness the interaction between the WA, as it expands outwards, and gas clouds at the outskirts of the BLR and/or the inner NLR. Furthermore, the prominent low-ionisation Fe\,II feature at 4126\,\AA\ seen in the SDSS spectrum fades in the DESI epoch. This trend could be interpreted as evidence for stronger ionisation in the DESI data, which would suppress Fe\,II emission. We note, however, that Fe\,II emission is extremely sensitive to local physical conditions such as density, optical depth, microturbulence and shielding. Therefore, the interpretation of the observed variations in this feature may be more complex \citep[e.g.][]{kovacevits2010}.

\subsection{Optical photometric properties}

The lack of strong temporal variations of $g$ band flux or the inferred optical extinction of 2MASS\,J14302580+4159572 has implications for the obscurer. Either the ionised material is dust free or the angular size subtended by the structure is smaller than the region of the accretion disc that dominates the optical ($g$ band) emission. In the latter case, using accretion disc temperature profile models \citep[e.g.][]{Novikov_Thorne1973, Papadakis2022} we estimate an upper limit of a few light-days ($\rm \lessapprox 10^{16}\, cm$) for the angular size of the ionised absorber.

\subsection{Comparison with WAs in the local Universe}

\cite{laha2014} present results from the characterisation of WAs along the line of sight in a sample of 26 X-ray unobscured AGN. They report that warm absorption is detected in more than 50\% of their sample, with ionisation parameter  $\log\,\xi$ in the interval [--1, 3.5]. The column density  is found to be in the range $10^{20}$ to $5\times10^{22}\rm \,cm^{-2}$, while outflow velocities from a few hundred to thousands km/s are estimated,  see also \citep{mckernan2007}.  Moreover, they report a positive correlation between the ionisation parameter and both the outflow velocity and the column density of the WA. Our assumed outflow (escape) velocity is consistent with their best fit relation. However, the column density measured by our analysis lies at the high-end of their distribution. Their best-fit line predicts a column density of at most a few times $10^{21}$ cm$^{-2}$ for $\log\,\xi$=1.5 (see right panel of their Fig. 5), which is much lower than the inferred column density for 2MASS\,J14302580+4159572. \cite{laha2014} also report a paucity of sources in the interval $\log\,\xi$=0.5-1.5, which is interpreted as evidence for thermal instabilities operating in this regime and efficiently dismantling clouds. It is interesting that the inferred ionisation parameter of 2MASS\,J14302580+4159572 overlaps with the thermally unstable zone proposed by \cite{laha2014}. The disagreement suggests that this zone may be less pronounced particularly for denser clouds ($N_H\gtrsim 10^{22}\rm \,cm^{-2}$) similar to those measured for 2MASS\,J14302580+4159572. We nevertheless caution that modelling assumptions may also be responsible for the differences between our analysis and the work of \cite{laha2014}.

\subsection{Comparison with thermally-driven ionised wind models}

Next we compare the inferred properties of the ionised clouds in 2MASS\,J14302580+4159572 with the predictions of models that link X-ray WAs with thermally-driven winds. This scenario, originally proposed by \citet{begelman1983}, describes how disc material is heated to the Compton temperature by reprocessing X-rays from the corona. The heated gas then expands because of the resulting pressure gradient, ultimately producing a thermally driven wind at the Compton radius, i.e. the distance at which the sound speed exceeds the local escape velocity  \citep[see also][]{doro2008, Mizumoto2019, laha2021}. \citet{Mizumoto2019} present quantitative models for thermal winds and predict the launch radius, velocity, column density and ionisation state as a function of black hole mass and accretion rate. The comparison of our results with these models is limited by observational uncertainties in the estimation of the black hole mass (and hence Eddington ratio) of  2MASS\, J14302580+4159572 as well as lack of constraints on the geometry of the system. Assuming $M_{BH}=1.7\times10^8$ $\rm M_\odot$  and $\log\lambda_{EDD}\simeq-1$ \citep{wu2022} for the source and  following \citet{Mizumoto2019} (see their Fig. 5), we obtain a rough wind launch radius estimate of $\sim 2\times10^{18}$\,cm, consistent with our findings.  Their predictions for the wind velocity, of the order of a few hundred kilometres per second, are also in broad agreement with our assumptions, particularly when the associated uncertainties are taken into account. In contrast, their model predictions for the ionisation state and column density of the wind (see their Figs. 10 and 11) are in tension with our findings. Specifically, our estimates of the ionisation parameter are significantly lower than the model predictions, while our measured column densities are considerably higher. This discrepancy cannot be attributed to the observational uncertainties. Nevertheless, \citet{Mizumoto2019} note that for higher black hole masses ($\sim 10^8$ solar masses), where the AGN outflow launch radius shifts outwards, their model may underpredict the column density because of the presence of additional material and differences in geometry. Since 2MASS\,J14302580+4159572 is itself a high‑mass AGN, this effect is likely relevant in our case, and part of the discrepancy may therefore arise from this limitation of the model.

\section{Conclusions}

Our multi-epoch X-ray and optical analysis of 2MASS\,J14302580+4159572 offers a rare, detailed portrait of the life cycle of a transient ionised obscuration event in a moderately distant type-I QSO. The source transitioned from an unobscured state in 2002--2003 to an absorbed phase in 2005, driven by a moderately ionised WA with $\rm N_{H} \sim 5\times10^{22}\,cm^{-2}$ and $\rm log\,\xi \approx 1.4$. Longer-term monitoring with \textit{Swift}/XRT shows that the absorber remained stable for nearly 18~years before weakening in 2023-2024, while the intrinsic 3--10~keV continuum remained constant throughout the entire 22-year baseline. 

The obscurer itself appears to reside well outside the BLR, at sub-parsec to parsec distances, with a density of $10^{6}$~cm$^{-3}$. Its influence appears to extend beyond the X-ray band. There is evidence for subtle variations in the [O\,III]~$\lambda4363$\,\AA\ and Fe\,II\,$\lambda$4126\,\AA\ optical emission lines between the 2003 SDSS and 2022 DESI spectra, suggesting a gradual modification of the inner narrow-line region on decadal timescales.

Our analysis favours the interpretation of the observed phenomenon as a warm, ionised outflow crossing our line of sight. Quantitative models for thermally driven winds are in tension with our findings. The observationally inferred ionisation parameters are significantly lower than predicted, while the measured column densities are considerably higher. This system thus stands as a compelling case study of the structure, longevity, and large-scale impact of ionised absorbers in quasars beyond the local Universe, and highlights the unique power of long-baseline archival monitoring for capturing the full lifecycle of such phenomena.

\begin{acknowledgements}
The authors acknowledge support from the Hellenic Foundation for Research and Innovation (HFRI) project "4MOVE-U"grant agreement 2688, which is part of the programme "2nd Call for HFRI Research Projects to support Faculty Members and Researchers" and the EU HORIZON-MSCA-2023-DN Project 101168906 "TALES: Time-domain Analysis to study the Life-cycle and Evolution of Supermassive black holes". This research is based on observations obtained with \xmm, an ESA science mission with instruments and contributions directly funded by ESA Member States and NASA. This research has made use of data from the \nustar\  mission, a project led by the California Institute of Technology, managed by the Jet Propulsion Laboratory, and funded by the National Aeronautics and Space Administration. Data analysis was performed using the \nustar\  Data Analysis Software ({\sc NuSTARDAS} ), jointly developed by the ASI Science Data Center (SSDC, Italy) and the California Institute of Technology (USA). This research has made use of data obtained from the \textit{Chandra} Data Archive and the \textit{Chandra} Source Catalog, and software provided by the \textit{Chandra} X-ray Center (CXC) in the application packages CIAO and Sherpa.  This research has made use of data and software provided by the High Energy Astrophysics Science Archive Research Center (HEASARC), which is a service of the Astrophysics Science Division at NASA/GSFC. This research uses data supplied by the UK \swift\ Science Data Centre at the University of Leicester. 
\end{acknowledgements}

\bibliography{WarmQSO}
\bibliographystyle{aa}

\begin{appendix}

\section{SED fitting}\label{sec:sed}

\begin{figure}

\includegraphics[width=0.48\textwidth]{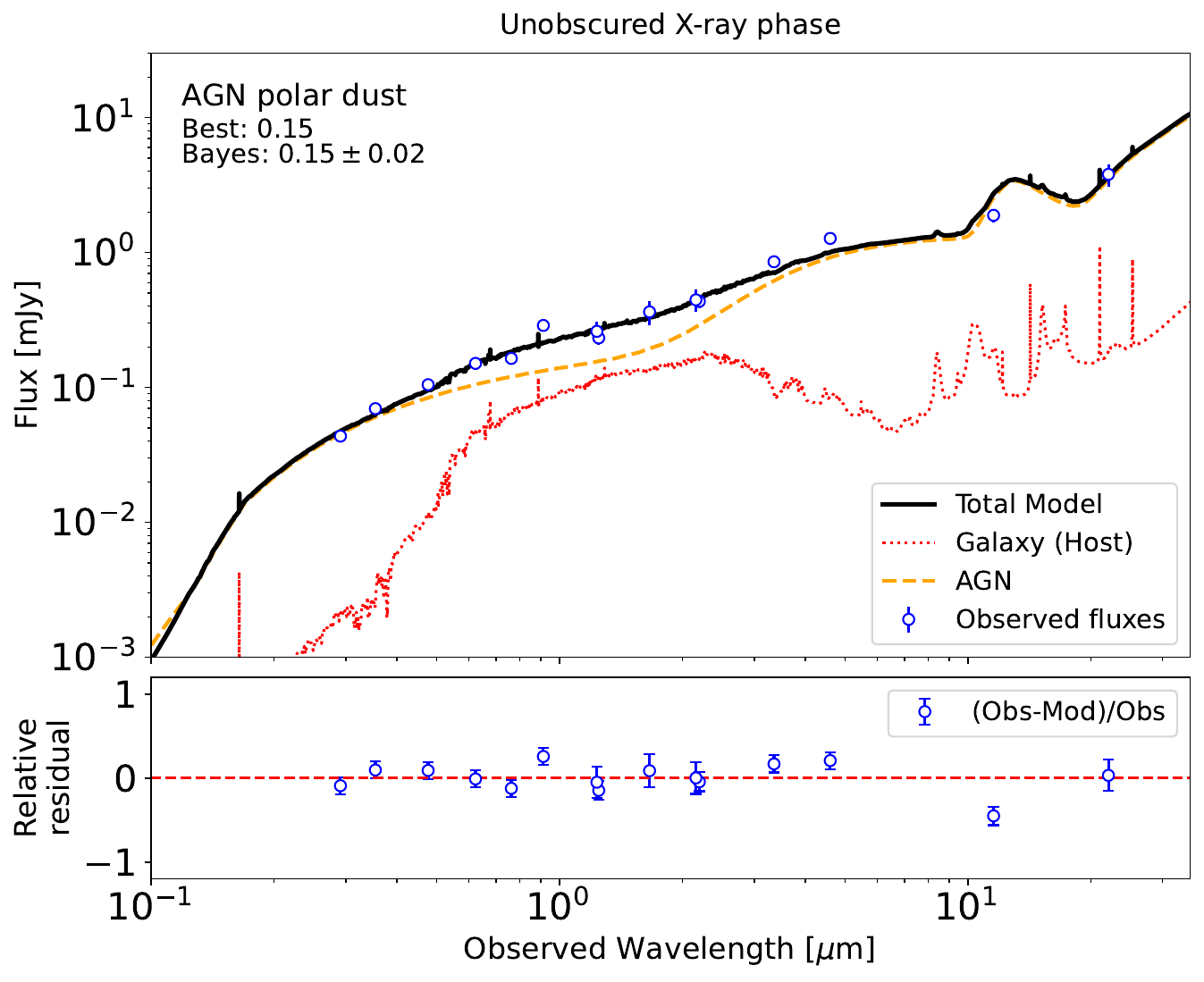} \\ 
\includegraphics[width=0.48\textwidth]{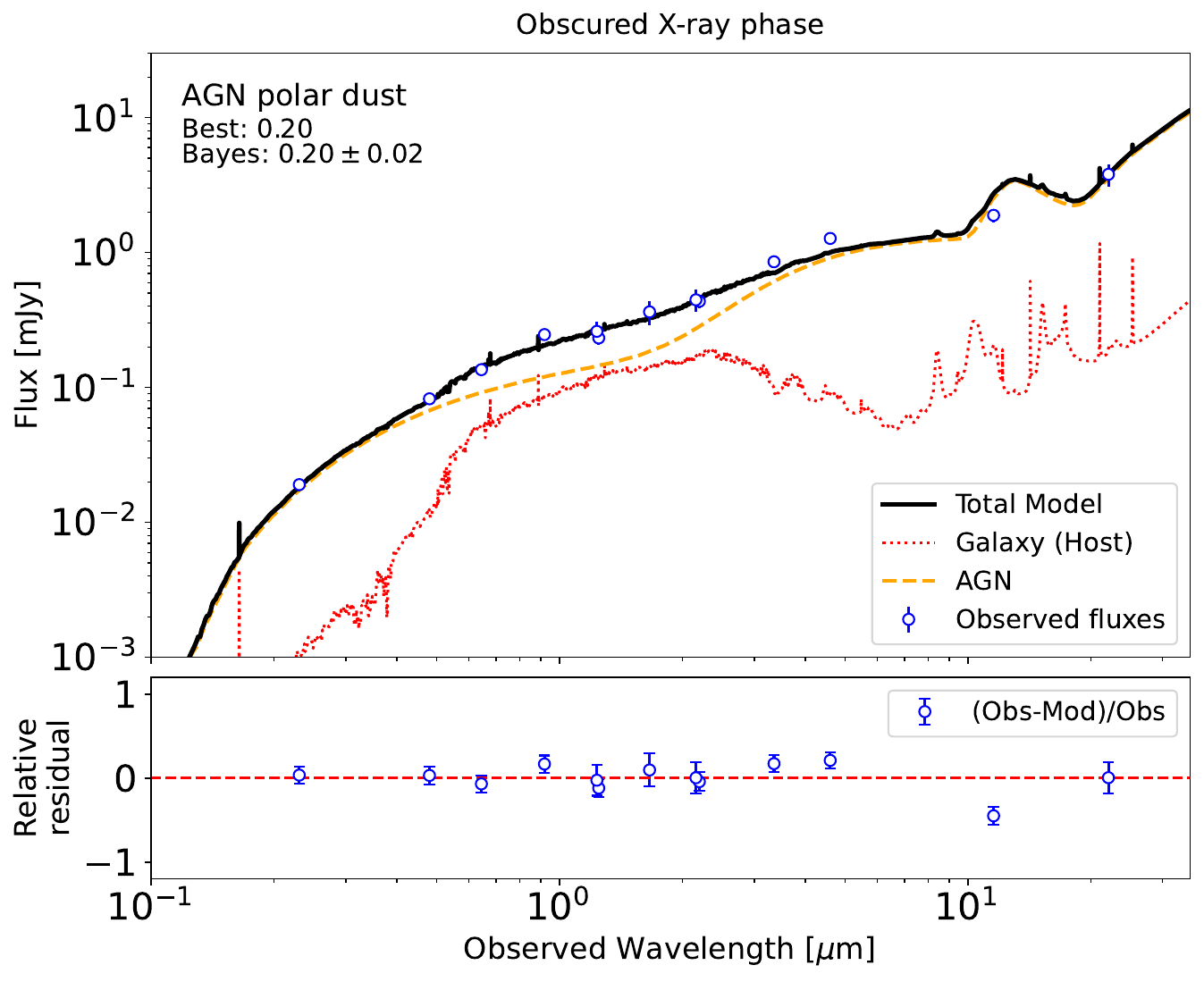}  

\caption{Spectral energy distribution fitting of 2MASS\,J14302580+4159572 at two distinct epochs. The top panel corresponds to the unobscured phase, constrained by \XMM/OM and SDSS photometry. The bottom panel shows the obscured phase, constrained by GALEX and LS10 photometry. In both panels, the solid black line represents the total best-fit model, the red dotted line indicates the attenuated stellar emission from the host galaxy, and the orange dashed line is the AGN contribution. The observed data points with associated errors are shown as blue circles. The lower sub-panels display the relative residuals, defined as $(F_{obs}-F_{mod})/F_{obs}$.}\label{plot_sed} 
\end{figure}

To characterise the physical properties of the source during its two distinct stages of activity, we fit templates to the observed SED using CIGALE \citep{yang2020, yang2022}. Following the results of the X-ray analysis, which reveal a prolonged interval of strong X-ray obscuration lasting approximately from 2005 to 2023, we divided the optical/UV data into two groups. The first includes measurements obtained during the unobscured X-ray phase, while the second consists of those obtained within the obscured X-ray phase. This distinction is useful for comparing the optical properties of the source with its X-ray behaviour, especially since no simultaneous optical–X-ray observations are available. We model the unobscured and obscured phases as two independent fitting runs to investigate potential variations in the total SED morphology and to determine whether changes in the observed flux are driven by modifications in the circumnuclear obscuration.

For both epochs, we use a common baseline of near-infrared (2MASS, UKIDSS) and mid-infrared (WISE) photometry to constrain the host galaxy and the dust torus emission. The time-variable components were handled as follows: for the unobscured epoch, we use SDSS optical photometry combined with \XMM\ OM (UVW1) data, corresponding to the period when the AGN was identified as unobscured in X-rays. For the obscured epoch, we use Legacy Survey DR10 optical data and GALEX NUV photometry observed on 2005-06-19 with the Deep Imaging Survey \citep[DIS,][]{bianchi2017}, which align with the phase of X-ray obscuration (obscured phase).

Our model grid follows the configuration described in \citet{Pouliasis2025}. Given the spectroscopic identification of the source as a broad-line QSO, we fix the inclination angle to \(i = 30^\circ\). Nuclear extinction is incorporated through an AGN polar dust grid with \(E(B-V)\) values of 0.0, 0.05, 0.1, 0.15, 0.2, and 0.3. All remaining components—including the stellar population models \citep{Bruzual2003}, the SKIRTOR AGNs templates \citep{Stalevski2012,Stalevski2016}, and the dust-law prescription  \citep{Charlot2000}—are kept consistent with the baseline of \citet{Pouliasis2025}.

In Fig.~\ref{plot_sed}, we present the two SEDs corresponding  to the unobscured/obscured X-ray phases. In both panels, the total model is shown in black, the attenuated emission from the host galaxy in red dotted lines, and the AGN component in orange dashed lines. Below each SED, we show the relative residual fluxes as a function of observed wavelength. Both fits yielded a reduced $\chi^2 \approx 1.7$. 

\end{appendix}

\end{document}